\documentclass[10pt]{article}

\usepackage[style=ieee, backend=biber]{biblatex}
\AtBeginBibliography{\footnotesize}

\usepackage{times}
\usepackage{graphicx}
\usepackage[normalem]{ulem}
\usepackage{adjustbox}
\usepackage{booktabs}
\usepackage{makecell}
\usepackage{multirow}

\usepackage[switch]{lineno}
 
\usepackage{float, subcaption} 

\usepackage{color}
\usepackage[dvipsnames]{xcolor}
\usepackage{hyperref}
\hypersetup{
	colorlinks,
	linkcolor={red!100!black},
	citecolor={blue!100!black},
	urlcolor={blue!80!black}
}
\usepackage{amssymb, bm}
\usepackage{amsmath}
\usepackage{microtype}
\usepackage[font={footnotesize}]{caption}
\usepackage{nomencl}
\usepackage[labelfont=bf]{caption}
\usepackage{cleveref}
\usepackage{mathrsfs}

\usepackage[indent=10pt]{parskip}
\usepackage{tocloft}

\usepackage{titlesec}
\titleformat{\section}
  {\normalfont\fontsize{12}{15}\bfseries}{\thesection}{1em}{}
\titleformat{\subsection}
  {\normalfont\fontsize{11}{12}\bfseries}{\thesubsection}{1em}{}

\titlespacing*{\subsubsection}{0pt}{\parskip}{4pt} 
\titlespacing\section{0pt}{12pt plus 4pt minus 10pt}{6pt plus 4pt minus 4pt}
\titlespacing\subsection{0pt}{3.0ex plus 4pt minus 0pt}{6pt plus 4pt minus 4pt}

\titleformat{\subsubsection}[runin]{\normalfont\bfseries}{\thesubsubsection.}{3pt}{}[.]   

\usepackage[total={7.0in, 9.5in}]{geometry}

\let\oldequation\equation
\let\oldendequation\endequation
\renewenvironment{equation}
  {\linenomathNonumbers\oldequation}
  {\oldendequation\endlinenomath}

\title{Accelerated, physics-inspired inference of skeletal muscle microstructure from diffusion-weighted MRI}

\author{Noel Naughton$^{1\star}$, Stacey Cahoon$^{2}$, Brad Sutton$^{1}$, and John G. Georgiadis$^{2}$\\[6pt] 
\normalsize{\textsuperscript{1}Beckman Institute for Advanced Science and Technology, University of Illinois at Urbana-Champaign, Urbana, IL.} \\
\normalsize{\textsuperscript{2}Department of Biomedical Engineering, Illinois Institute of Technology, Chicago IL.} 
\normalsize{\textsuperscript{$\star$}nnaught2@illinois.edu} 
}

\date{}

\begin{document}

\twocolumn[
  \begin{@twocolumnfalse}
    \maketitle
    \begin{abstract}
    Muscle health is a critical component of overall health and quality of life. However, current measures of skeletal muscle health take limited account of microstructural variations within muscle, which play a crucial role in mediating muscle function.
    To address this, we present a physics-inspired, machine learning-based framework for the non-invasive and  \textit{in vivo} estimation of microstructural organization in skeletal muscle from diffusion-weighted MRI (dMRI). 
    To reduce the computational expense associated with direct numerical simulations of dMRI physics, a polynomial meta-model is developed that accurately represents the input/output relationships of a high-fidelity numerical model. 
    This meta-model is used to develop a Gaussian process (GP) model to provide voxel-wise estimates and confidence intervals of microstructure organization in skeletal muscle. Given noise-free data, the GP model accurately estimates microstructural parameters.
    In the presence of noise, the diameter, intracellular diffusion coefficient, and membrane permeability are accurately estimated with narrow confidence intervals, while volume fraction and extracellular diffusion coefficient are poorly estimated and exhibit wide confidence intervals. 
    A reduced-acquisition GP model, consisting of one-third the diffusion-encoding measurements, is shown to predict parameters with similar accuracy to the original model. 
    The fiber diameter and volume fraction estimated by the reduced GP model is validated via histology, with both parameters within their associated confidence intervals, demonstrating the capability of the proposed framework as a promising non-invasive tool for assessing skeletal muscle health and function.
    \end{abstract}
    \vspace{15pt}
  \end{@twocolumnfalse}
]

\section{Introduction} 
Muscle health is strongly correlated to quality of life \cite{Kell2001, Samuel2011}, motivating a clinical need for interventional methods and quantitative diagnostics that improve muscle health.
This need is particularly acute for aging populations, as age-related loss of muscle mass is a primary determinant of decreased muscle function and mobility \cite{Roubenoff2001}, both of which are linked to increased mortality \cite{Villareal2004}.
Skeletal muscle exhibits a hierarchical structure of elongated, tightly-packed muscle fibers that are surrounded by multiple levels of collagenous extracellular matrix, which plays an important role in force transmission \cite{purslow2002structure}.
However, traditional measures of muscle health take limited account for these structural features, restricting our understanding of muscle's structure--function relationship \cite{Correa-de-Araujo2017a}. 
Non-invasive measurements of skeletal muscle structure is thus positioned to enable novel insights into the physiological changes of muscle during aging \cite{Heymsfield2015} or muscle pathology \cite{Wokke2014}, aiding the development of effective, targeted treatments to increase muscle health.

Currently, biopsy and histology is the most common modality to quantitatively investigate skeletal muscle microstructure, but this measurement approach is invasive, labor-intensive, and highly local to the excised muscle region. 
A promising alternative with potential to address these limitations is diffusion-weighted magnetic resonance imaging (dMRI), which can provide \textit{in-vivo}, non-invasive characterization of muscle microstructure organization over the entire muscle volume. 
dMRI is sensitive to the direction-dependent diffusion distance of water in tissue. In muscle, water diffuses faster in a muscle fiber's axial direction than transverse direction, where barriers such as cell walls restrict diffusion, resulting in anisotropic diffusion behavior that sensitizes the voxel's MR signal to these $\sim$$\mu$m length-scale structures.
Thus, although the resolution limit of clinical dMRI is $\sim$mm, the underlying tissue microstructure is encoded in the MR signal of each voxel.

\begin{figure*}[t!]
    \centering
    \includegraphics[width=\textwidth]{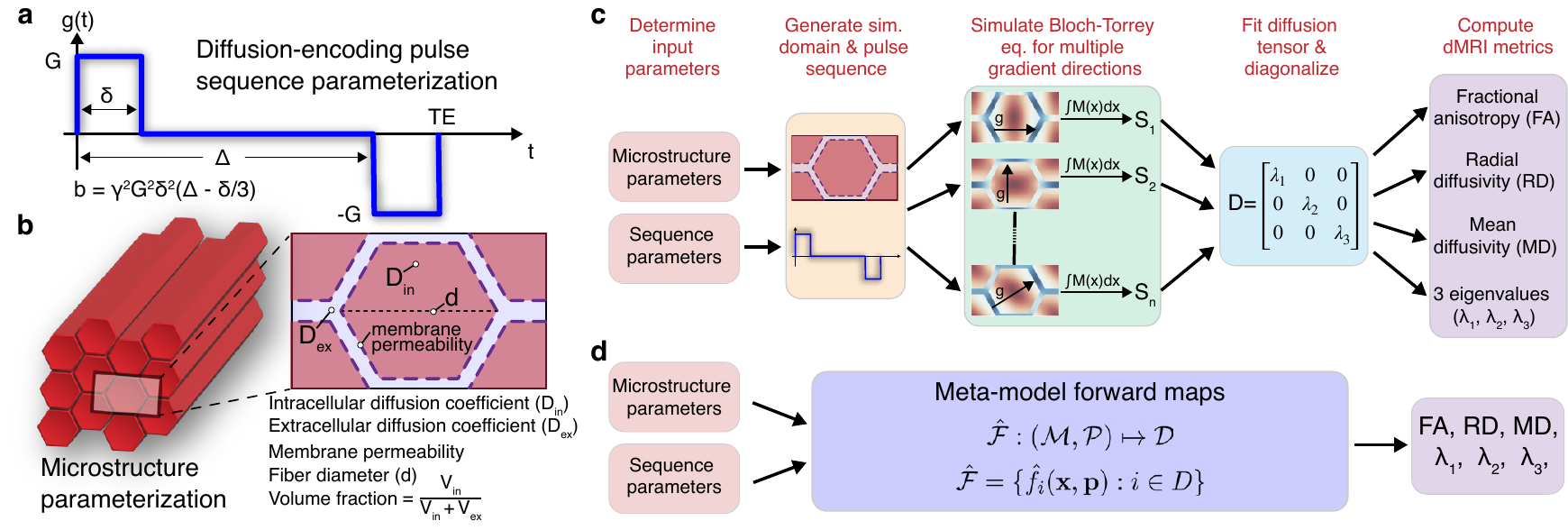}
    \caption{\textbf{\textit{In-silico} experiments of skeletal muscle dMRI.} (a) Schematic of the generalized diffusion sequence used to represent the diffusion-encoding sequence. (b) Schematic of numerical model of skeletal muscle as periodically packed hexagonal cylinders. (c) Overview of numerical simulation process with the generation of a pulse sequence and microstructural domain, the independent simulation of multiple diffusion directions, construction of diffusion tensors, and finally estimation of dMRI metrics. (d) Overview of meta-model approach where the microstructural and pulse sequence parameters are directly mapped to the dMRI metrics. Here standard DTI metrics are used, but more complex diffusion models can be straightforwardly included.}
    \label{fig:1}
\end{figure*}

Decoding the relationship between skeletal muscle microstructure and dMRI measurements is, however, not straightforward. 
While the forward problem of how microstructure variations influence the dMRI signal has been extensively characterized using numerical models \cite{Naughton2019MRM,Berry2018,Berry:2021,Bates2017}, the inverse problem of estimating microstructural parameters from dMRI remains unresolved. 
Though accurate, numerical models are in general too computationally expensive to use as the basis of solving the inverse problem \cite{hwang2003image,Xu2009,beltrachini2015parametric,Nguyen2018,Hall2017, Berry2018,Naughton2019LBM_arxiv}, and, while simplified compartmental models have been used in the past \cite{Saotome2006, Laghi2017, Karampinos2009,Galban2005,Kim2005}, such models do not accurately capture diffusion behavior in skeletal muscle \cite{Naughton2019PMB}. 
Another recent model, the Random Permeable Barrier Model, abstracts muscle as a domain of random barriers \cite{Fieremans2017,Winters2018} but does not account for differences in the intracellular and extracellular domain, which is necessary if changes in the extracellular matrix, a critical mediator of skeletal muscle function \cite{purslow2002structure}, are to be considered.
Thus, there remains a need for models that can accurately characterize the microstructural organization of skeletal muscle.

Here we combine the accuracy and realism of physics-based numerical models with the computational speed of analytical and data-driven models to develop a framework for estimating the microstructural organization of skeletal muscle from dMRI measurements. 
Beginning with a numerical model of dMRI, we develop a computationally efficient method of inversion while maintaining the accuracy of the numerical solution. 
We first develop an accelerated solution of the forward problem before using this solution as the basis of a data-driven inverse solution. We conclude with an experimental demonstration of its viability. Overall, this work provides a flexible framework for development of physics-inspired inversion models for the non-invasive estimation of tissue microstructure. While applied here to skeletal muscle modeled using a simplified periodic domain, this methodology is broadly applicable to many classes of biological tissues such as neural and cancer tissues.

\section{Fast evaluation of dMRI metrics in muscle }

To understand how a voxel's diffusion MRI (dMRI) signal can be used to estimate the underlying microstructural organization of skeletal muscle, we first consider the forward problem of how microstructural variations influence the dMRI signal and formulate a data-driven, accelerated solution procedure that will aid in the development of an inverse mapping solution. 

\subsection{Numerical simulation and parameterization}\label{sec:parameterization}

\subsubsection*{Bloch-Torrey equation} dMRI physics is governed by the Bloch-Torrey equation \cite{torrey1956bloch}, which describes the time evolution of the dMRI signal in a voxel as
\begin{equation}
\label{LBM-B-Teq}
\frac{\partial \bm{M}}{\partial t} = -i \gamma \left( \bm{x} \cdot \bm{g}(t) \right) \bm{M} + \nabla \cdot \left(D(\bm{x}) \nabla \bm{M} \right)
\end{equation}  
where $\bm{M}(\bm{x},t)$ is the complex-valued, transverse spin magnetization resulting from exciting longitudinal spins onto the transverse plane and which is manipulated by an externally applied magnetic field ($\bm{x} \cdot \bm{g}(t)$). Here $\gamma$ is the gyromagnetic ratio of hydrogen, $\bm{x}$ is the spin position vector, $\bm{g}(t)$ is the time-varying magnetic field gradient vector used to encode diffusion, and $D(\bm{x})$ is the local diffusion coefficient. 

\subsubsection*{Pulse sequence parameterization} The magnetic field gradient $\bm{g}(t)$ is operator-controlled and can be manipulated to probe different aspects of tissue structure \cite{Drobnjak2010}. Here we focus on two related diffusion-encoding pulse sequences: the Stejskal-Tanner pulsed-gradient spin echo (PGSE) \cite{stejskal1965spin} and the simulated echo acquisition mode (STEAM) diffusion sequence \cite{Tanner1970}. If $T_1$ and $T_2$ effects are set aside, both sequences can be minimally described by the generalized diffusion-weighted sequence described in Fig. \ref{fig:1}a that consists of a bipolar magnetic gradient pulse of magnitude $G$ \cite{Rose2019,Naughton2019MRM}.
This generalized sequence is parameterized by four variables, echo time (TE), gradient duration ($\delta$), gradient spacing ($\Delta$), and b-value ($b=\gamma^2 G^2 \delta^2 (\Delta - \delta/3)$). Here, gradient duration is fixed at 5 ms and TE is defined as TE = $\Delta + \delta$, resulting in two free variables of gradient spacing and b-value. The diffusion time of the spins is $t_{d} = \Delta - \delta/3$. However, because our fixed gradient duration of 5 ms is short relative to the gradient spacing timings considered here, for simplicity we approximate the gradient spacing as the diffusion time (i.e. $t_{d} = \Delta$). 

\subsubsection*{Muscle tissue model parameterization} 
At the microstructural level, skeletal muscle consists of parallel, elongated fibers, each surrounded by a semi-permeable membrane (sarcolemma) and embedded in an extracellular matrix. 
Informed by histologically based simulations \cite{Naughton2019LBM_arxiv}, we represent muscle's microstructural organization via a compact domain of infinitely long, parallel hexagonal cylinders arranged in a periodic array (Fig.~\ref{fig:1}b). We define a representative elemental volume (REV), which we parameterize to provide a parsimonious description of the muscle microstructure
consisting of two morphological parameters (fiber diameter and muscle fiber volume fraction) and three tissue parameters (membrane permeability and intra-/extracellular diffusion coefficients). 
Water diffusion in the intra- and extracellular domains is characterized by homogeneous (effective) diffusion coefficients that capture the cumulative effects of sub-cellular restrictions within each domain. 

\subsubsection*{Numerical simulations} 
The Bloch-Torrey equation of Eq.~\ref{LBM-B-Teq} is integrated using the lattice Boltzmann method (LBM) on a D3Q7 stencil, supplemented with appropriate boundary conditions and the initial condition $\bm{M}(\bm{x},t=0) = 1$.
Intra-domain, semi-permeable boundary conditions handle the effect of spins crossing the muscle's sarcolemma membrane while modified periodic boundary conditions on the domain boundaries accommodate the non-periodic magnetization accumulation over the periodic REV geometry. 
Full details of both the boundary conditions and the numerical lattice Boltzmann scheme implementation are available in Naughton et al. \cite{Naughton2019LBM_arxiv}

Solving the Bloch-Torrey equation over the prescribed domain results in a spatially localized distribution of the MR signal $\bm{M}(\bm{x},t=\text{TE})$(see Fig. \ref{fig:1}c). However, in a physical MR experiment, image formation integrates this local signal distribution to provide the voxel's dMRI signal.
Performing a similar integration to the numerical simulation result
\begin{equation}
\label{integrate_signal}
S = \left|\left|\int_{V}\bm{M}(\bm{x},t=\text{TE})d\bm{x}\right|\right|
\end{equation}
where $V$ is the voxel volume allows matching the numerical simulation result to the MR signal that would be measured on a scanner. 
This equivalence effectively allows the performance of \textit{in silico} dMRI experiments, where a known microstructural domain can be defined and a dMRI experiment performed to numerically evaluate the signal such a domain would yield if measured \textit{in vivo}. 

A schematic overview of the simulation pipeline for these \textit{in silico} experiments is given in Fig.~\ref{fig:1}c.
For each experiment, six non-collinear gradient directions ($S_i$) and a non-diffusion-weighted acquisition ($S_0$) are simulated and used to fit a diffusion tensor using the fanDTasia ToolBox \cite{fanDTasia}. 
In muscle, the diffusion tensor is anisotropic and described by three eigenvalues ($\lambda_1$, $\lambda_2$, and $\lambda_3$), which correspond to the tensor's principal directions. These eigenvalues are then used to compute the tensor invariants of fractional anisotropy
$\text{FA} = \sqrt{\frac{(\lambda_1-\lambda_2)^2 +(\lambda_2- \lambda_3)^2+ (\lambda_3- \lambda_1)^2}{2(\lambda_1^2+ \lambda_2^2+ \lambda_3^2)}}$,
mean diffusivity MD = $(\lambda_1+ \lambda_2+ \lambda_3)/3$,
and radial diffusivity RD = $(\lambda_2+ \lambda_3)/2$, 
which characterize the diffusion anisotropy and magnitude within the voxel. 
While here a standard diffusion tensor imaging (DTI) model is used, more complex post-processing and diffusion models can straightforwardly be incorporated.

\subsection{Model acceleration via meta-modeling}
While accurate, numerical simulation of the forward problem is computationally expensive, with a typical \textit{in silico} dMRI experiment taking 1-2 minutes per voxel. Coupled with direct inverse solution approaches, which require many iterative forward solutions, use of this numerical model results in solution times on the order of 45 minutes per voxel to estimate underlying microstructural parameters \cite{PEARC2019}. Scaled to a typical 64$\times$64 (or larger) resolution dMRI image with multiple slices, this approach quickly becomes computationally infeasible.

To accelerate solution times, we exploit the insight that the spatial distribution of the signal within the microstructural domain---which is computationally expensive to obtain---is not directly needed and can be bypassed by the deployment of a data-driven forward mapping 
\begin{equation}
    \mathcal{F} : (\mathcal{M}, \mathcal{P}) \mapsto \mathcal{D}
\end{equation} that directly maps the microstructural ($\mathcal{M}\in \mathbb{R}^m$) and diffusion-encoding sequence ($\mathcal{P} \in \mathbb{R}^p$) domains to the dMRI metrics ($\mathcal{D} \in \mathbb{R}^d$). 

\subsubsection*{Polynomial meta-model} 
We approximate the forward mapping $\mathcal{F}$ using a set of meta-models to generate individual mappings for each dMRI metric 
\begin{equation}
    \hat{\mathcal{F}}=\{\hat{f}_i(\mathbf{x}, \mathbf{p}) : i\in D\}
\end{equation}
for $D \in \{$FA, MD, RD, $\lambda_1$, $\lambda_2$, $\lambda_3$$\}$. These meta-models (Fig.~\ref{fig:1}d) approximate the relationship between the inputs ($m=5$ microstructural parameters $\mathbf{x} \in \mathcal{M}$ and $p=2$ sequence parameters $\mathbf{p} \in \mathcal{P}$) and outputs ($d=6$ dMRI metrics) of the numerical model pipeline in a data-driven manner with no regard for the underlying physics. 

While a number of meta-modeling frameworks \cite{Barton2006} and machine learning techniques \cite{Karniadakis2021,Wang2021} are available, here we adopt the polynomial chaos expansion approach, which models the evolution of a system with stochastic inputs \cite{PolyChaos-Eldred2009, PolyChaos-Kaintura2018}. 
Each meta-model is represented as an expansion of a properly selected polynomial basis truncated to a finite basis set
\begin{equation}
\hat{f}_i(\mathbf{x}, \mathbf{p}) = \sum_{j=0}^T \alpha_{j} \ \bm{\psi}_{j} (\mathbf{z})
\end{equation}
where $\alpha_j$ are the basis weights, $\bm{\psi}_j$ are multivariate polynomials, $T+1 = (N+C)!/(N!~C!)$ is the number of terms in the basis set for a maximum polynomial order of $C$ and $N=m+p$ input parameters, and $\mathbf{z} = \Gamma(\mathbf{x}, \mathbf{p})\in [-1,1]^N$ is the normalized transformation of the uniformly distributed microstructural parameter ranges given in Table \ref{table:fitting_parameters} to the interval $[-1,~1]$.
The choice of basis polynomials $\bm{\psi}_j$ is determined by the parameter distribution, which, for the uniform distribution considered here, is the set of Legendre polynomials \cite{Xiu2002}.

\begin{table}[t!]
    \begin{center}
        \caption{Microstructure and pulse sequence parameter ranges spanned by meta-model. }
        \label{table:fitting_parameters}
        \label{table:parameter_range_naiveLBM}
        \begin{tabular}
            {l r c r l }
            \toprule
            Parameter & \multicolumn{3}{c}{Range} \\
            \midrule    
            Muscle fiber diameter   & 10  &--& 80   & $\mu$m        \\
            Volume fraction         & 0.7 &--& 0.95 &               \\
            Membrane permeability   & 10  &--& 100  & $\mu$m/s    \\
            Intracellular diffusion & 0.5 &--& 2.5  & $\mu$m$^2$/ms \\
            Extracellular diffusion & 0.5 &--& 2.5  & $\mu$m$^2$/ms \\
   \midrule
        Diffusion time  & 10 &--& 750  & ms \\ 
        b-value & 300 &--& 1200  & s/mm$^2$ \\
            \bottomrule
        \end{tabular}
    \end{center}
    \vspace{-15pt}
\end{table}
 
\begin{figure*}[t!]
    \centering
    \includegraphics[width=\textwidth]{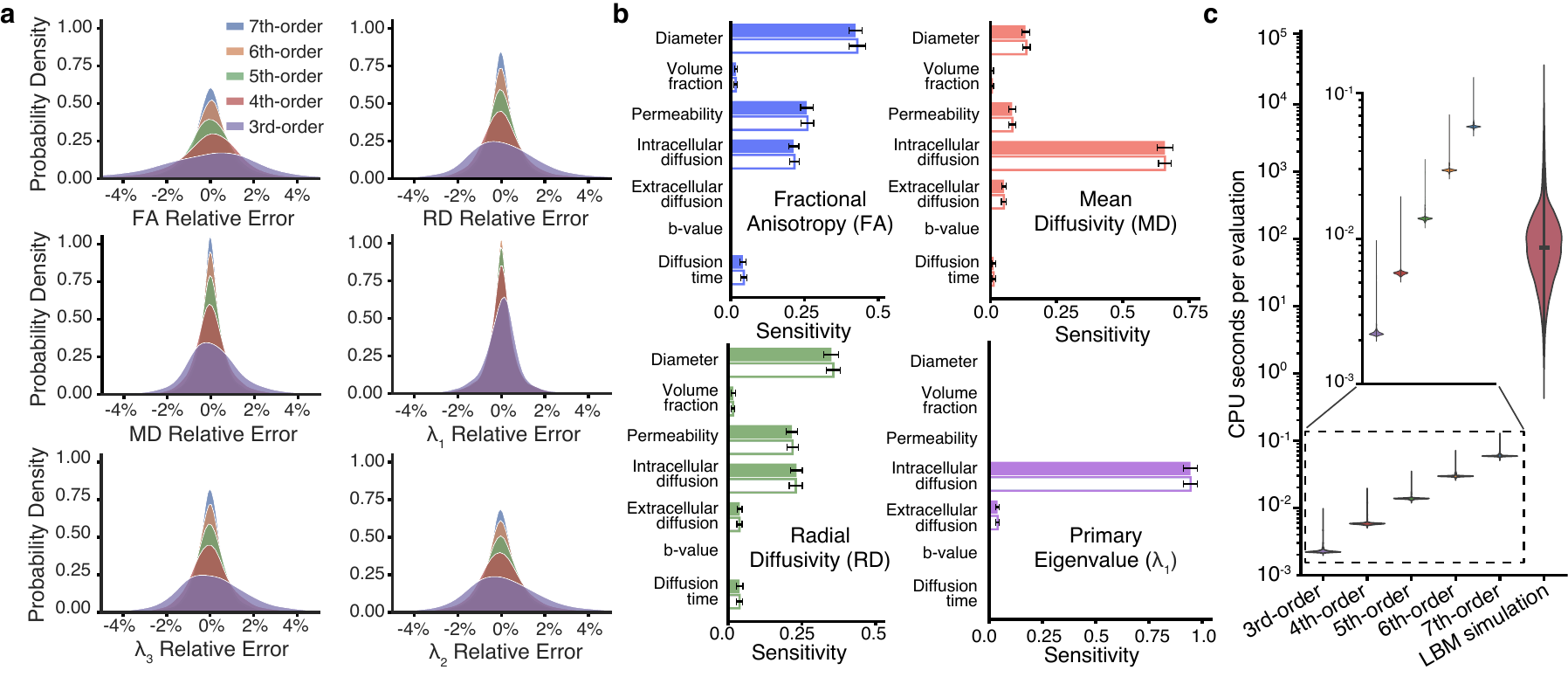}
    \caption{\textbf{Meta-model acceleration of dMRI simulations.} (a) Density plots of meta-model error for dMRI metrics of FA, MD, RD, $\lambda_1$, $\lambda_2$, and $\lambda_3$. As the order of the polynomial basis set increases, the meta-model accuracy improves. (b) Global sensitivity indices of the 7$^\text{th}$-order meta-model (solid color) and LBM numerical model (white) demonstrating the meta-model captures the global behavior of parameter variation. Results for $\lambda_2$ and $\lambda_3$ and not shown as they are nearly identical to RD results. (c) The computational cost of meta-model evaluation increases with polynomial order but remains lower than the numerical model in all cases. The 7$^\text{th}$-order meta-model has a mean evaluation time 3 orders of magnitude (1000x) faster than the numerical model and a maximum evaluation time 5 orders of magnitude faster.  }
    \label{fig:2}
\end{figure*}

\subsubsection*{Meta-model generation} 
To generate the meta-model, a previously reported  data set of 80,000 parameter sets 
\cite{Naughton2019MRM} was used to fit the weights $\alpha_j$ of the PC expansion using a 70/30 train/test split. Data was fit using least-squares linear regression by the open-source \textit{ChaosPy} Python package \cite{chaospy} .
The dataset was generated via a Sobol sampling method, which is a low discrepancy sampling method, meaning the input data used to construct the model are evenly distributed across the input space \cite{Sobol1967-sequence}. 
Each dataset entry consists of a microstructural parameter set $\mathbf{x}$, a pulse sequence parameter set $\mathbf{p}$, and  dMRI metrics $f_i(\mathbf{x},\mathbf{p})$ for $i\in D$ that resulted from an \textit{in silico} dMRI experiment following the procedure described in Section \ref{sec:parameterization}. 
Maximum polynomial orders $C \in \{3, ... 7\}$ of the Legendre polynomial basis were considered. 

\subsubsection*{Accuracy and sensitivity} 
Evaluating the trained meta-models over the test data resulted in a vector of meta-model estimates $\mathbf{\hat{f}}_i$ and an associated vector of ground-truth, numerically-simulated dMRI metrics $\mathbf{f}_i$.
To quantify the accuracy of the meta-model we consider the accuracy of each dMRI metric independently and define a relative error metric 
\begin{equation}
    \text{e}_i = \frac{\hat{f}_i(\mathbf{x},\mathbf{p}) - f_i(\mathbf{x},\mathbf{p})}{\max(\textbf{f}_i) - \bar{\textbf{f}_i}}
\end{equation}
where $\bar{\textbf{f}_i}$ is the mean of the vector $\textbf{f}_i$. 

Figure~\ref{fig:2}a shows the distribution of relative error between the meta-models and the ground-truth numerical simulations in the test split of the data set for each dMRI metric. For increasing polynomial order, the accuracy of the meta-model improves, with errors on the order of $<$1\%  for the meta-model constructed from 7$^\text{th}$-order polynomials, showing it to be an accurate representation of the underlying numerical model.
The test data used to quantify accuracy is extracted from the same Sobol sequence as the training data, implying these results are an upper limit of the meta-model error as the test data is sampled at points far away from the training data. 
To further quantify the performance of the meta-model, the sensitivity of the meta-model to changes in its inputs was computed and compared to the sensitivity of the underlying numerical model reported in \cite{Naughton2019MRM}.
The results, shown in Fig. \ref{fig:2}b, show the sensitivity indices are nearly identical to those of the full numerical model, demonstrating that the fitted meta-model accurately captures the global responses of the underlying numerical model for each microstructural input parameter.

\subsubsection*{Computational efficiency} 
As the polynomial order increases, the accuracy of the meta-model increases, but so too does the computational evaluation time (Fig.~\ref{fig:2}c). However, the trained meta-models are drastically faster to evaluate than the underlying LBM numerical model for all polynomial orders, with the mean evaluation time of the 7$^\text{th}$-order meta-model three orders of magnitude faster than the mean evaluation time of the numerical model. This speed-up becomes even more pronounced when maximum evaluation time is considered, with the 7$^\text{th}$-order meta-model five orders of magnitude faster than the numerical model. The time-stepping nature of the LBM numerical model leads to its evaluation time being directly proportional to the diffusion time of the simulated sequence. The analytical nature of the meta-model, in contrast, obviates this limitation, leading to more uniform evaluation times. 

\section{Inverse problem solution}
We next turn to the inverse problem of estimating microstructural organization from dMRI data. Previous attempts have generally consisted of iteratively solving a forward model to converge on a set of microstructural parameters \cite{PEARC2019,Saotome2006, Laghi2017, Karampinos2009,Galban2005,Kim2005,Fieremans2017}; however, such approaches are computationally expensive due to the large number of function evaluations required. 
Our goal here is to instead define a data-driven inverse map \begin{equation}
    \mathcal{G} : \mathcal{S} \mapsto \mathcal{M}
\end{equation} that directly maps the dMRI measurement domain $\mathcal{S} \in \mathbb{R}^{s \times d}$ to the microstructure parameter domain $\mathcal{M}\in \mathbb{R}^m$. 
While machine learning regression approaches have been widely used to generate cost-effective inverse solutions for a range of problems \cite{Karniadakis2021, Arridge2019, Ongie2020}, many traditional approaches, such as random forest regression or deep neural networks, do not provide uncertainty measures for their outputs, which can be critically important to interpreting results. To address this, here we utilize Gaussian process (GP) regression, or kriging, to generate a data-driven inverse map that additionally provides confidence intervals of its predictions, substantially increasing the interpretability, and thus utility, of the estimates \cite{Rasmussen2004}.

\subsection{Gaussian process regression}
Each microstructural parameter is modeled as its own Gaussian process, leading to the inverse map
\begin{equation}
    G=\{g_i(\mathbf{s}) : i\in M\}    
\end{equation}
where $\mathbf{s}\in S$ is a vector of dMRI metrics acquired by a set of $s$ diffusion-encoding sequences (defined below) and $M$ is the domain of all five microstructural parameters. 

\begin{figure*}[t!]
    \centering
    \includegraphics[width=\textwidth]{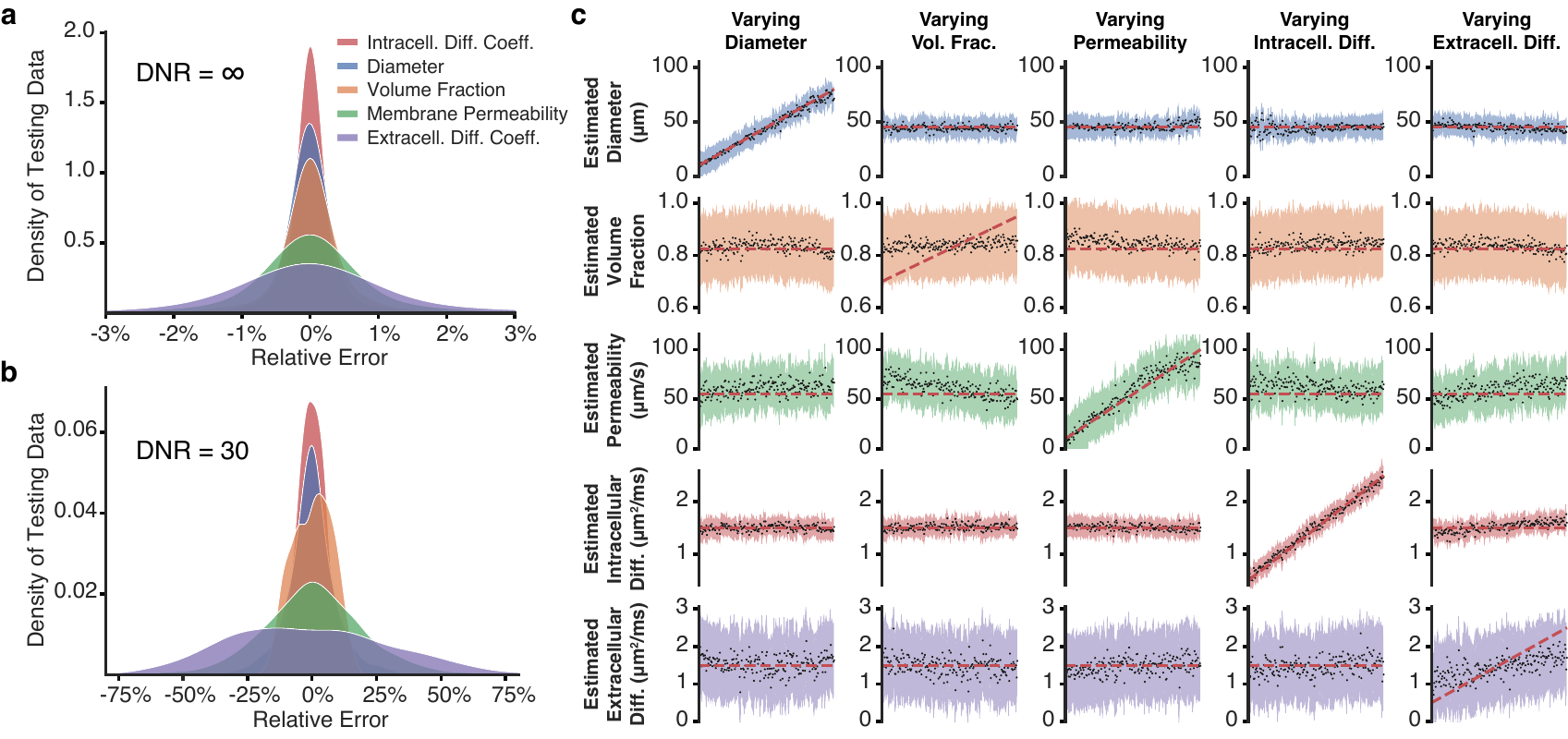}
    \caption{\textbf{Gaussian process inverse mapping.} (a) Error density of GP model when evaluating synthetic, noise-free dMRI measurements. (b) Error density of GP model when evaluating noisy dMRI measurements with a diffusion-to-noise ratio of 30. (c) GP model results for dMRI measurements (DNR=30) when a single microstructural parameter is varied at a time. For each column, 200 meta-model evaluations were made as the varied microstructure parameter was linearly increased. Results are organized column-wise with vertical alignment relating to the same dMRI measurement and the dashed line denoting the true microstructural value. }
    \label{fig:3}
\end{figure*}

A Gaussian process is a generalization of a Gaussian probability distribution. It is a collection of random variables, any finite subset of which has a joint Gaussian distribution \cite{GPbook-Williams2006}. A Gaussian process
\begin{equation}
g_i(\mathbf{s}) \sim \mathcal{GP}\left(m_i(\mathbf{s}),k_i(\mathbf{s},\mathbf{s}')\right) {.}
\end{equation}
is defined by a mean function $m_i(\mathbf{s})$ and covariance function $k_i(\mathbf{s},\mathbf{s}')$ over the inputs $\mathbf{s}$. 
In practice, is it common to either subtract out the mean or directly set $m_i(\mathbf{s}) = 0$, allowing the Gaussian process to be written as
\begin{equation}
g_i(\mathbf{s}) \sim \mathcal{N}\left(0,k_i(\mathbf{s},\mathbf{s}')\right)
\end{equation}
where $\mathcal{N}(\cdot)$ denotes a normal distribution. 

A key component of Gaussian process regression is the selection of the covariance kernel function $k_i(\mathbf{s},\mathbf{s}')$, which can strongly influence the model's accuracy. 
Progress has been made towards automated kernel selection \cite{Duvenaud2013}; however empirical kernel selection is still generally required. 
Here, a radial basis function kernel and a linear kernel are combined with a Gaussian noise kernel to form the covariance function
\begin{equation}
\label{eq:kernel}
\begin{split}
k_i(\mathbf{s},\mathbf{s}') =&  \sigma^2_r \exp\left(-\frac{1}{2l^2}(\mathbf{s} - \mathbf{s}')^2\right) + \\ 
&\sigma^2_l(\mathbf{s} - c)(\mathbf{s}'-c) + \sigma^2_n \delta_{ij}
\end{split}
\end{equation}
where kernel variances $\sigma$ and length-scales $l$, $c$ are hyperparameters tuned to maximize the log marginal likelihood of the model over the training data, which consists of the set of microstructural parameters $\mathbf{g}_i$ and associated dMRI measurements $S = \{\mathbf{s}_1, \mathbf{s}_2,...,\mathbf{s}_n\}$.
A unique set of hyperparameters is tuned for each GP $g_i(\mathbf{s})$.
While it is possible to incorporate information about relationships between microstructural parameters (coregionalization), such methods substantially increase the difficulty of hyperparameter optimization and often lead to over-fitting  \cite{GPbook-Williams2006}. As such, they were not considered here.

With the kernel defined, the inverse problem can be formulated as:
given a set of $n$ training observations of microstructural parameters $\mathbf{g}_i$ and their set of corresponding dMRI measurements $S$, predict the microstructural parameter distribution $\mathbf{g}_i^* \sim \mathcal{N}(0,K_i(S^*,S^*))$ given the set of $n^*$ dMRI measurements $S^*$. 
According to the definition of Gaussian processes, the joint probability distribution of these training and evaluation outputs is also Gaussian and can be written
\begin{equation}
\label{eq:joint_dist}
\begin{bmatrix} \mathbf{g}_i \\ \mathbf{g}_i^* \end{bmatrix}
\sim \mathcal{N} \left(0, 
\begin{bmatrix}
K_i(S,S) & K_i(S,S^*) \\
K_i(S^*,S) & K_i(S^*,S^*)
\end{bmatrix}
\right) 
\end{equation}
where, $K_i(S^*,S)$ is the $n^* \times n$ matrix of the covariances of $S^*$ and $S$ based on Eq. \ref{eq:kernel} (and similarly for all $K_i(\cdot,\cdot)$).

To include the training data information in our estimation of $\mathbf{g}_i^*$, we condition the joint Gaussian prior distribution of Eq.~\ref{eq:joint_dist} on the observations $\mathbf{g}_i$, resulting in the posterior distribution
\begin{equation}
\mathbf{g}_i^*|S^*,S,\mathbf{g}_i  \sim  \mathcal{N}\left(\bar{\mathbf{g}}_i^*,\mathrm{cov}(\mathbf{g}_i^*)\right)
\end{equation}
where the mean $\bar{\mathbf{g}}_i^*$ and covariance matrix $\mathrm{cov}(\mathbf{g}_i^*)$ are
\begin{equation}
\label{GP_mean}
\bar{\mathbf{g}}_i^*  =  K_i(S^*,S)K_i(S,S)^{-1}\mathbf{g}_i
\end{equation}
\begin{equation}
\label{GP_cov}
\mathrm{cov}(\mathbf{g}_i^*)  =  K_i(S^*,S^*) - K_i(S^*,S) 
K_i(S,S)^{-1}K_i(S,S^*)
\end{equation}
Note that both the mean $\bar{\mathbf{g}}_i^*$ and covariance matrix $\mathrm{cov}(\mathbf{g}_i^*)$ are solely functions of the training observations $\mathbf{g}_i$, the training dMRI data $S$, and the observed dMRI data $S^*$, all of which are known. 
Thus, given a set of dMRI measurements $\mathbf{s}$ , for each microstructural parameter we can evaluate the predicted posterior distribution to compute a mean microstructural estimate $\mu_i(\mathbf{s})$ (Eq. \ref{GP_mean}) and its variance $\sigma_i(\mathbf{s})$ (Eq. \ref{GP_cov}), which is then scaled to determine a 95\% confidence interval.

\subsection{dMRI-inversion model formulation}\label{inversion_preliminaries}
To formulate an inverse mapping, a fixed set of diffusion-encoding parameter sets must first be defined. 
Because multiple microstructural parameters are being estimated, multiple diffusion-encoding measurements are required to constrain the inverse problem. 
Here, we focus on the effect of diffusion time and b-value.
Multiple b-values allows sensitivity to non-Gaussianity of the diffusion behavior \cite{Grebenkov2010} while varying diffusion time sensitizes the diffusion behavior to microstructural features at different length scales \cite{Fieremans2017, Porcari2018, Porcari2020}.

Five b-values (400, 600, 800, 1000, and 1200 s/mm\textsuperscript{2}) and six diffusion times (20, 50, 100, 200, 400, 700 ms) were selected to generate $s=30$ unique diffusion-encoding parameter sets.
All sequences are here modeled as STEAM sequences represented by the generalized diffusion sequence with described in Section \ref{sec:parameterization}.
The b-value is achieved by adjusting the diffusion-encoding gradient strength.
Both b-values and diffusion times were selected to span the ranges viable for a clinical scanner using a STEAM sequence for skeletal muscle dMRI.

\subsubsection*{Model training} To generate training data for the GP model, the 7$^\text{th}$-order meta-model was sampled using a Sobol sequence at 9000 microstructural parameter combinations. The meta-model was evaluated for FA, $\lambda_1$, and RD for all thirty diffusion-encoding parameter sets, resulting in 90 input measurements ($\lambda_2$ and $\lambda_3$ were not considered due to their strong similarity to RD). To reduce the dimensionality of the input data, these 90 inputs were combined into a single vector for each parameter set, and principal component analysis (PCA) was performed on the vectors. The 20 dimensions that best described the observed input variance were selected, scaled to zero-mean and unit-variance, and used as the inputs to the GP model.

The meta-model produces noise-free estimates of dMRI metrics; however, accounting for the influence of noise in dMRI measurements is a critical consideration of any inverse solution \cite{Damon2008} and so synthetic noise was injected into the training data. Because the meta-model directly outputs dMRI metrics, it is not possible to incorporate noise directly into the raw signal. Instead, Gaussian noise is injected into the dMRI metrics based on a diffusion-to-noise (DNR) ratio. Five copies of the dMRI metrics were created. One remained noise-free with noise added to the others to achieve a DNR of 30. 

The open-source \textit{GPy} Python package was used to fit the GP model and optimize the GP kernel hyperparameters using the L-BFGS-B algorithm \cite{gpy2014}.  
For large data sets such as those considered here, this optimization can be slow. To accelerate the process, subsets of increasing size containing only noise-free data were iteratively used to optimize the hyperparameters over smaller data sets, allowing the GP model to quickly learn the coarse structure of the underlying data. 

\begin{figure*}[t!]
    \centering
    \includegraphics[width=\textwidth]{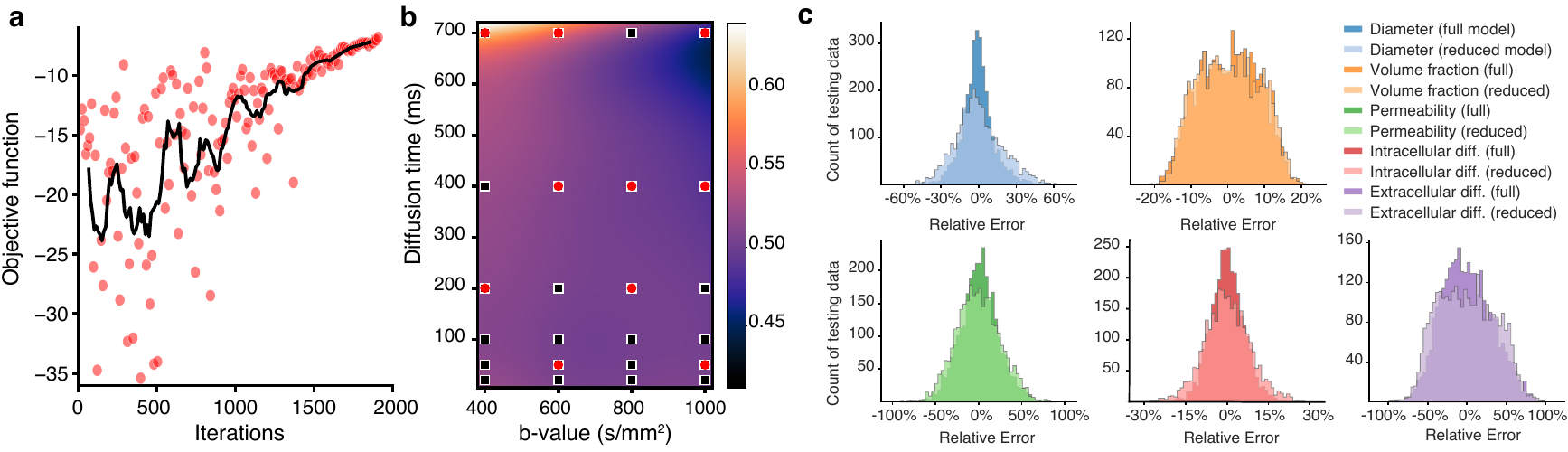}
    \caption{\textbf{Optimized diffusion-encoding sequence selection.} (a) Convergence of CMA-ES algorithm's maximization of the objective function over 2000 iterations to select a compact set of ten diffusion-encoding sequences. (b) Map of the variance between the 1024 different microstructural parameters used for different diffusion time and b-value combinations. The locations of the sequence used for the full GP model (black boxes; b=1200 s/mm2 not visualized) and ten sequences selected by the CMA-ES algorithm (red circles) are overlaid. (c) Comparisons of the relative error histograms of the full and reduced sequence GP models for all microstructural parameters shows limited loss in accuracy when the reduced set of sequences are used. }
    \label{fig:4}
\end{figure*}

\subsubsection*{GP model accuracy}
The 7$^\text{th}$-order meta-model was sampled using a new Sobol sequence to generate 3000 data-points. Evaluating the test data resulted in vectors of mean microstructural estimates $\bm{\mu}_i$ for $i \in M$ and associated vectors of ground-truth microstructural values $\mathbf{x}_i$. To quantify GP model accuracy, we define the relative error metric
\begin{equation}
    \text{e} = \frac{\mu_i(\mathbf{s}) - x_i(\mathbf{s})}{\bar{\mathbf{x}}_i}
\end{equation}
where $x_i(\mathbf{s})$ is the true microstructural parameter.
The accuracy of the model was quantified against both noise-free (Fig.~\ref{fig:3}a) and noisy (DNR = 30; Fig.~\ref{fig:3}b) versions of the test data. While the accuracy of microstructural estimates decreases in the presence of noise, the GP model's high accuracy when given noise-free data demonstrates both the general invertability of the problem and that the GP model is learning the underlying data structure.  

A third test is depicted in Fig.~\ref{fig:3}c wherein the test data is generated by evaluating the forward meta-model and varying only a single 
microstructural parameter at a time (diagonal entries) while holding all others constant. 
As Fig.~\ref{fig:3}c shows, the GP model is able to accurately capture changes in fiber diameter, membrane permeability, and intracellular diffusion coefficient while struggling to identify variations in the volume fraction or extracellular diffusion coefficient. 
However, the GP model's confidence intervals in its mean estimate of the volume fraction and extracellular diffusion coefficient are wide, indicating the mean parameter estimate should be interpreted with caution. In contrast, its accurate estimates of fiber diameter, membrane permeability, and intracellular diffusion coefficient are accompanied by comparably narrower confidence intervals, indicated confidence in the mean estimate.

\subsection{Reduced diffusion-encoding model}\label{reduced_pulse} 
While it is broadly known how different diffusion-encoding pulse profiles affect the MR signal \cite{Grebenkov2010}, which combinations of sequences encode the most microstructural information about the tissue is less clear. In the prior section, this uncertainty is addressed by densely sampling the possible combinations of diffusion time and b-value (Fig.~\ref{fig:4}b). However, because imaging time is proportional to the total number of diffusion-encoding parameter sets utilized, reducing the number of parameter sets used is paramount for clinical translatablility. 

To identify a compact set of diffusion-encoding sequences, the 5$^\text{th}$-order meta-model (for computational efficiency) was sampled at four uniformly spaced points for all five microstructural parameters ($N=1024$) and FA, MD, RD, and $\lambda_1$ metrics were precomputed for 80 pulse sequences defined on a grid of four evenly-spaced b-values between 400 and 1000 s/mm\textsuperscript{2} and twenty evenly-spaced diffusion times between 10 and 700 ms. 
For a candidate subset of $s=10$ sequences, a 3$^{\text{rd}}$-order polynomial surface was fit to the dMRI metrics at each of the $N$ microstructural parameter sets. The fitted surface was evaluated at all 80 pulse sequences resulting in a vector $\mathbf{q}_{i_n}$ for $i\in \{ \text{FA, MD, RD, $\lambda_1$} \}$ that was compared to the precomputed meta-model dMRI metrics $\mathbf{Q}_{i_n}$. An objective function was defined to quantify the ability of a subset of ten diffusion-encoding pulse sequences to capture the structural information encoded by this full set of sequences according to
\begin{equation}
    \label{eq:obj_fn}
   \phi = -\sum_{n=1}^{N} \Big|\Big| \big( || \mathbf{q}_{i_n} - \mathbf{Q}_{i_n} || : i\in \{ \text{FA, MD, RD, $\lambda_1$} \} \big) \Big|\Big|
\end{equation}
We maximized Eq.~\ref{eq:obj_fn} using the CMA-ES algorithim (Fig.~\ref{fig:4}a), which is a non-linear, derivative-free evolutionary search algorithm \cite{Hansen2003CMA-ES}. The ten sequences selected by the CMA-ES algorithm were matched to the nearest corresponding pulse sequences used to define the previous GP model (Fig.~\ref{fig:4}b) 

\begin{figure*}[t!]
    \centering
    \includegraphics[width=\textwidth]{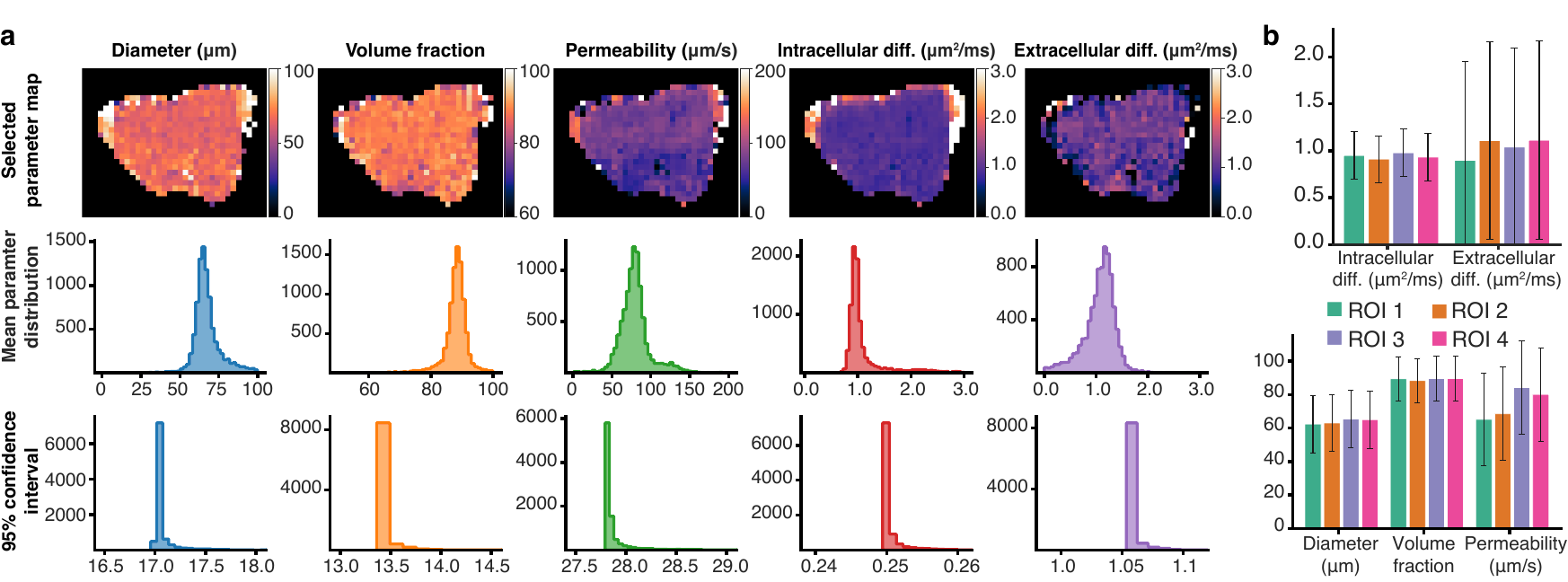}
    \caption{\textbf{Experimental estimates of microstructure organization.} (a) Voxel-wise estimates of all five microstructural parameters for a single slice of the bovine \textit{biceps femoris} dMRI data (top row); histograms of the distribution of the five microstructural parameters over the entire 3D domain after thresholding to exclude outliers related to edge voxels (middle row); and distribution of 95\% confidence intervals from GRP model over entire 3D domain (bottom row). (b) Averaged estimates and 95\% confidence intervals for all five microstructural parameters of four ROIs within the tissue.  }
    \label{fig:5}
\end{figure*}

\subsubsection*{Reduced GP model}
A second GP model was fit using this reduced set sequences following the same process as before. A comparison of the full and reduced GP model's accuracy is shown in Fig.~\ref{fig:4}c, where, to test the model's generalization performance, both models are tested against a dataset of 3000 solutions generated by the LBM numerical model. Results for both models are broadly similar, with only a slight decrease in accuracy for the reduced pulse sequence model. Using only one-third the diffusion-encoding measurements, this set of $s=10$ sequences defines a more experimentally plausible set of sequences for use in a physical experiment. 

\section{Experiments}

\subsection{Voxel-wise microstructure estimation}
To test the accuracy of the proposed inversion scheme, a validation experiment was performed. Tissue from a bovine \textit{biceps femoris} was procured and a 7 cm sample from a unipenate portion was excised and imaged on a Siemens Prisma 3T scanner using a prototype diffusion-weighted STEAM sequence with EPI readout \cite{Tanner1970}. Ten diffusion-weighted measurements based on the sequences selected by the CMA-ES algorithm were obtained, each with twelve gradient directions and 2 averages. Twenty slices with a $64\times64$ FOV matrix and 3.4 mm isotropic voxels were acquired. For all scans, TE = 49 ms. For scans with $\Delta > 500$ ms, TR = 1500 ms, otherwise TR = 1200 ms.  
The experiment took approximately one hour to complete. Image data was thresholded and post-processed with FSL \cite{fsl} to compute diffusion metrics for each voxel. This dataset (9607 voxels) was read into the reduced GP model and used to produce voxel-wise estimates of the microstructural parameters. Here, $\lambda_3$ was used instead of RD to better capture microstructural data based on the hypothesis that $\lambda_3$ is a stronger reflection of fiber-level transverse diffusion behavior \cite{Galban2004,ISMRM_fascicle_ellipticity}. Additionally, four ROIs were manually defined within the 3D tissue volume, consisting of between 12 and 32 voxels. Diffusion metrics from these voxels were averaged and used to estimate mean microstructural parameters and confidence intervals for the ROI. 

Figure \ref{fig:5}a (top row) shows the resulting voxel-wise estimates of all five microstructural parameters for a representative slice of the 3D volume. For each microstructural parameter, estimation for all 9607 voxels took $39.18 \pm 0.05$ seconds (4 ms/voxel). Parameter ranges appear consistent throughout the muscle beyond slight variation in voxels near the edge of the tissue, where partial volume effects likely distorted the signal. Figure \ref{fig:5}b (middle row) shows histograms of the distribution of the five microstructural parameters after thresholding to exclude outliers related to edge voxels while Fig. \ref{fig:5}c (bottom row) shows the distribution of the associated 95\% confidence intervals. Mean parameter estimates have unimodal distributions with relatively small tails while the confidence intervals exhibit a sharp minimum bound, indicative of experimental noise limiting the confidence of the inversion scheme. Results from the four ROIs are shown in Fig.~\ref{fig:5}d. 

\subsection{Histological validation}
Validation of microstructure estimates is challenging due to the difficulty of independently measuring many of the estimated microstructural parameters, in particular intracellular and extracellular diffusion coefficients or sarcolemma membrane permeability. Morphological parameters, such as fiber diameter and volume fraction, are comparably easier to verify through histological examination. 
Tissue from two ROIs (1 and 2) was excised and fixed in 10\% buffered formalin for one week. After fixation, tissue was dehydrated, embedded in paraffin wax, sectioned, and stained with hematoxylin and eosin. Microscopy images were acquired at 40x magnification. 
Images were processed using ImageJ \cite{imageJ} to binarize data, and fiber diameters for each ROI were approximated by the median Feret diameter using the `Analyze Particles' tool in ImageJ. 
During tissue processing, distortion due to tissue shrinkage can skew measurements, possibly leading to underestimation of fiber diameter. 
To mitigate this bias and establish an upper bound of fiber diameters, binarized images were processed in Matlab with a watershed transform to skeletonize the domain and eroded using ImageJ to reestablish a uniformly thin extracellular space. This processing effectively swells the fibers to fill the domain and the Feret diameter was again measured. Muscle fiber volume fractions were also computed for both the original and swelled domains.

\begin{table*}[t!]
    \centering
    \caption{Diameter and volume fraction measurements from histological examination of ROIs 1 and 2 for the original and swelled domains as well as the GP model estimates with 95\% confidence interval (CI).  }
    \label{table:ROI_estimates}
    \begin{tabular}{ccccccc}
        \toprule
        &\multicolumn{3}{c}{Fiber diameter} &\multicolumn{3}{c}{Volume fraction} \\
        \hphantom{-}& Original & Swelled & GP estimate $\pm$ 95\% CI
                    & Original & Swelled & GP estimate $\pm$ 95\% CI\\
        \cmidrule(lr){2-4} \cmidrule(lr){5-7}   
        ROI 1 & 55.2 $\mu$m & 73.9 $\mu$m & 62.3 $\pm$ 17.0 $\mu$m  & 0.674 & 0.894 & 0.894 $\pm$ 0.134 \\
        ROI 2 & 51.0 $\mu$m & 62.0 $\mu$m & 62.9 $\pm$ 17.0 $\mu$m & 0.514 & 0.902 & 0.883 $\pm$ 0.134 \\
        \bottomrule
    \end{tabular}
\end{table*}

The histologically-measured fiber diameters and volume fractions of the ROIs are reported in Table \ref{table:ROI_estimates} along with the estimated mean and confidence intervals of the GP model.
The measured fiber diameters fall within the predicted range, though there is a relatively wide confidence interval.
The confidence intervals of the ROIs are lower than the voxel-wise results due to the averaging of the dMRI metrics over the ROI increasing the SNR of the data.
These results demonstrate that the GP model can provide estimates of the fiber diameter that agree with histological measurements. 
The swelled measurements of volume fraction also match the GP estimates while the results from the original histology measurements are lower than the estimated range. The large confidence range of the volume fraction cautions against giving too much credence to this parameter estimate and shrinkage and other deformation of the tissue during fixation and processing make it difficult to know which volume fraction estimate is the most valid.

\section{Discussion}
This paper presents a framework for computationally efficient estimation of skeletal muscle microstructural parameters from dMRI. 
Use of Gaussian processes (GP) in the inverse mapping provides not only voxel-wise microstructure estimates but also uncertainty intervals, which increases the utility and interpretability of the model's estimates by identifying when a microstructure estimate can be strongly relied on or when it is likely an arbitrary guess. There are two approaches to interpreting these confidence intervals. The first is adopted here, where a single mean value represents the microstructure of the entire voxel and the confidence intervals are then a measure of uncertainty. The second interpretation treats the predicted Gaussian distribution as representing a distribution of parameters within the voxel (e.g. distribution of fiber diameters). Considering the distribution of microstructures that occur in skeletal muscle, such a perspective may further extend the insights available from GP-based inverse models. 

In the presence of noise, the parameters the GP inversion scheme accurately estimates in Fig.~\ref{fig:3} (diameter, intracellular diffusion coefficient, and membrane permeability) are notably the parameters with the highest sensitivity indices of the forward models reported in Fig. \ref{fig:2}b. 
Given noise-free data, the inverse map model accurately inverts the problem for all five microstructural parameters, indicating the GP model learns the underlying data structure and suggesting model accuracy can be increased if higher SNR measurements are acquired.

The framework developed here has a flexible, modular structure. It consists of multiple independent components, each of which can be refined or exchanged with alternative approaches to improve future iterations. 
For example, incorporating more sophisticated parameterizations of muscle microstructure will further improve the realism of the forward problem, while use of alternative meta-modeling techniques, such as deep neural networks may increase the accuracy and computational efficiency of its meta-model approximation. 
The diffusion-encoding schemes considered here can also be advanced by increasing the fidelity of the PGSE and STEAM sequence simulations or by considering additional diffusion-encoding sequences such as OGSE \cite{Does:2003}. 
Finally, combining diffusion-encoding sequence optimization with compressed sensing frameworks may allow further reduction in the diffusion-encoding sequences needed, aiding in clinical feasibility and translatability efforts by further reducing imaging time.

\section{Conclusion}

Overall, this work provides a flexible, modular framework for development of physics-inspired, data-driven inversion models for the estimation of tissue microstructure from dMRI. 
To reduce the computational expense associated with direct numerical simulations of dMRI physics, a polynomial meta-model is developed that accurately represents the numerical model and is used to develop a Gaussian process regression model to provide voxel-wise microstructural estimates and confidence intervals.  
The proposed methodology is broadly applicable to additional classes of biological tissues such as neural and cancer tissues, extending its potential impact.
Applied here to skeletal muscle, its experimental implementation and validation demonstrates the capability of the framework as a promising non-invasive tool for \textit{in vivo} assessment of skeletal muscle health and function.

{\small
\vspace{2pt}
\subsubsection*{Data Availability} \vspace{-5pt}
Code and model weights for the meta-model and Gaussian process regression models presented in this paper are available upon request from the authors. \par}
 
{\small
\subsubsection*{Acknowledgements} \vspace{-5pt}
This work was supported in part by an NSF GRFP (N.N.), the R.A. Pritzker endowed chair (J.G.), NSF grants CMMI-1437113, and CMI-1762774 (J.G.), and NIH grants HL090455 and EB018107 (J.G.). \par
}
\vspace{-6pt}
  
{\footnotesize
\printbibliography

@article{Does:2003,
  title={Oscillating gradient measurements of water diffusion in normal and globally ischemic rat brain},
  author={Does, Mark D and Parsons, Edward C and Gore, John C},
  journal={Magnetic Resonance in Medicine},
  volume={49},
  number={2},
  pages={206--215},
  year={2003},
  publisher={Wiley Online Library}
}

@book{Rasmussen2004,
  title={Advanced Lectures on Machine Learning: ML Summer Schools 2003, Canberra, Australia, February 2-14, 2003, T{\"u}bingen, Germany, August 4-16, 2003, Revised Lectures},
  author={Bousquet, Olivier and von Luxburg, Ulrike and R{\"a}tsch, Gunnar},
  volume={3176},
  year={2011},
  publisher={Springer}
}

@article{Porcari2020,
  title={Time-dependent diffusion MRI as a probe of microstructural changes in a mouse model of Duchenne muscular dystrophy},
  author={Porcari, Paola and Hall, Matt G and Clark, Chris A and Greally, Elizabeth and Straub, Volker and Blamire, Andrew M},
  journal={NMR in Biomedicine},
  volume={33},
  number={5},
  pages={e4276},
  year={2020},
  publisher={Wiley Online Library}
}

@article{Berry:2021,
  title={Varying diffusion time to discriminate between simulated skeletal muscle injury models using stimulated echo diffusion tensor imaging},
  author={Berry, David B and Englund, Erin K and Galinsky, Vitaly and Frank, Lawrence R and Ward, Samuel R},
  journal={Magnetic resonance in medicine},
  volume={85},
  number={5},
  pages={2524--2536},
  year={2021},
  publisher={Wiley Online Library}
}

@inproceedings{Duvenaud2013,
  title={Structure discovery in nonparametric regression through compositional kernel search},
  author={Duvenaud, David and Lloyd, James and Grosse, Roger and Tenenbaum, Joshua and Zoubin, Ghahramani},
  booktitle={International Conference on Machine Learning},
  pages={1166--1174},
  year={2013},
  organization={PMLR}
}

@article{Ongie2020,
  title={Deep learning techniques for inverse problems in imaging},
  author={Ongie, Gregory and Jalal, Ajil and Metzler, Christopher A and Baraniuk, Richard G and Dimakis, Alexandros G and Willett, Rebecca},
  journal={IEEE Journal on Selected Areas in Information Theory},
  volume={1},
  number={1},
  pages={39--56},
  year={2020},
  publisher={IEEE}
}

@article{Arridge2019,
  title={Solving inverse problems using data-driven models},
  author={Arridge, Simon and Maass, Peter and {\"O}ktem, Ozan and Sch{\"o}nlieb, Carola-Bibiane},
  journal={Acta Numerica},
  volume={28},
  pages={1--174},
  year={2019},
  publisher={Cambridge University Press}
}

@article{Karniadakis2021,
  title={Physics-informed machine learning},
  author={Karniadakis, George Em and Kevrekidis, Ioannis G and Lu, Lu and Perdikaris, Paris and Wang, Sifan and Yang, Liu},
  journal={Nature Reviews Physics},
  volume={3},
  number={6},
  pages={422--440},
  year={2021},
  publisher={Nature Publishing Group UK London}
}

@article{Wang2021,
  title={Metamodeling of constitutive model using Gaussian process machine learning},
  author={Wang, Jikun and Li, Tianjiao and Cui, Fan and Hui, Chung-Yuen and Yeo, Jingjie and Zehnder, Alan T},
  journal={Journal of the Mechanics and Physics of Solids},
  volume={154},
  pages={104532},
  year={2021},
  publisher={Elsevier}
}

@article{Barton2006,
  title={Metamodel-based simulation optimization},
  author={Barton, Russell R and Meckesheimer, Martin},
  journal={Handbooks in operations research and management science},
  volume={13},
  pages={535--574},
  year={2006},
  publisher={Elsevier}
}

@InProceedings{ISMRM_fascicle_ellipticity,
  author    = {Naughton, Noel and Wang, Anthony and Georgiadis, John},
  title     = {Fascicle Ellipticity as an Explanation of Transverse Anisotropy in Diffusion {MRI} Measurements of Skeletal Muscle},
  booktitle = {Proceedings of the 27th Annual Meeting of the International Society for Magnetic Resonance in Medicine},
  address={Montreal, Canada},
  note={Abstract 1276},
  year      = {2019},
}

@article{Naughton2019MRM,
  title={Global sensitivity analysis of skeletal muscle dMRI metrics: Effects of microstructural and pulse parameters},
  author={Naughton, Noel M and Georgiadis, John G},
  journal={Magnetic Resonance in Medicine},
  volume={83},
  number={4},
  pages={1458--1470},
  year={2020},
  publisher={Wiley Online Library}
}

@inproceedings{PEARC2019,
  title={Connecting Diffusion {MRI} to Skeletal Muscle Microstructure: Leveraging Meta-Models and {GPU}-acceleration},
  author={Naughton, Noel and Georgiadis, John},
  booktitle={Proceedings of the Practice and Experience in Advanced Research Computing on Rise of the Machines (learning)},
  pages={7},
  year={2019},
  organization={ACM}
}

@Article{Naughton2019LBM_arxiv,
  title={Lattice {Boltzmann} method for simulation of diffusion magnetic resonance imaging physics in multiphase tissue models},
  author={Naughton, Noel M and Tennyson, Caroline G and Georgiadis, John G},
  journal={Physical Review E},
  volume={102},
  number={4},
  pages={043305},
  year={2020},
  publisher={APS}
}

@Article{Naughton2019PMB,
  author    = {Naughton, Noel and Georgiadis, John},
  title     = {Comparison of two-compartment exchange and continuum models of {dMRI} in skeletal muscle},
  journal   = {Physics in Medicine {\&} Biology},
  year      = {2019},
  volume    = {64},
  number    = {15},
  pages     = {155004},
  abstract  = {Clinical diffusion MRI (dMRI) is sensitive to micrometer scale spin displacements, but the image resolution is  ∼mm, so the biophysical interpretation of the signal relies on establishing appropriate subvoxel tissue models. A class of two-compartment exchange models originally proposed by Kärger have been used successfully in neural tissue dMRI. Their use to interpret the signal in skeletal muscle dMRI is challenging because myocyte diameters are comparable to the root-mean-square of spin displacement and their membrane permeability is high. A continuum tissue model consisting of the Bloch–Torrey equation integrated by a hybrid lattice Boltzmann scheme is used for comparison. The validity domain of a classical two-compartment tissue model is probed by comparing it with the prediction of the continuum model for a 2D unidirectional composite continuum model of myocytes embedded in a uniform matrix. This domain is described in terms of two dimensionless parameters inspired by mass transfer phenomena, the Fourier (F) and Biot (B) numbers. The two-compartment model is valid when  and , or when  and . The model becomes less appropriate for muscle dMRI as the cell diameter and volume fraction increase, with the primary source of error associated with modeling diffusion in the extracellular matrix.},
  %doi       = {10.1088/1361-6560/ab2aa6},
  publisher = {{IOP} Publishing},
  %url       = {https://%doi.org/10.1088%2F1361-6560%2Fab2aa6},
}

@article{fsl,
	title={{FSL}},
	author={Jenkinson, Mark and Beckmann, Christian F and Behrens, Timothy EJ and Woolrich, Mark W and Smith, Stephen M},
	journal={Neuroimage},
	volume={62},
	number={2},
	pages={782--790},
	year={2012},
	publisher={Elsevier}
}

@Misc{gpy2014,
  author =   {{GPy}},
  title =    {{GPy}: A {Gaussian} process framework in python},
  howpublished = {\url{http://github.com/SheffieldML/GPy}},
  year = {since 2012}
}

@book{GPbook-Williams2006,
  title={Gaussian processes for machine learning},
  author={Williams, Christopher KI and Rasmussen, Carl Edward},
  year={2006},
  publisher={MIT press Cambridge, MA}
}

@article{PolyChaos-Kaintura2018,
  title={Review of polynomial chaos-based methods for uncertainty quantification in modern integrated circuits},
  author={Kaintura, Arun and Dhaene, Tom and Spina, Domenico},
  journal={Electronics},
  volume={7},
  number={3},
  pages={30},
  year={2018},
  publisher={Multidisciplinary Digital Publishing Institute}
}

@inproceedings{PolyChaos-Eldred2009,
author = {Michael Eldred},
title = {Recent Advances in Non-Intrusive Polynomial Chaos and Stochastic Collocation Methods for Uncertainty Analysis and Design},
booktitle = {50th AIAA/ASME/ASCE/AHS/ASC Structures, Structural Dynamics, and Materials Conference},
year={2009},
%doi = {10.2514/6.2009-2274},
%url = {https://arc.aiaa.org/%doi/abs/10.2514/6.2009-2274 },
%eprint = {https://arc.aiaa.org/%doi/pdf/10.2514/6.2009-2274 }
}

@article{Hansen2003CMA-ES,
  title={Reducing the time complexity of the derandomized evolution strategy with covariance matrix adaptation ({CMA-ES}).},
  author={Hansen, N. and Muller, S.D. and Koumoutsakos, P.},
  journal={Evolutionary Computation},
  volume={11},
  number={1},
  pages={1--18},
  year={2003},
  publisher={MIT Press}
}

@Article{stejskal1965spin,
  author    = {Stejskal, Edward O and Tanner, John E},
  title     = {Spin diffusion measurements: spin echoes in the presence of a time-dependent field gradient},
  journal   = {The Journal of Chemical Physics},
  year      = {1965},
  volume    = {42},
  number    = {1},
  pages     = {288--292},
  publisher = {AIP Publishing},
}

@Article{Xu2009,
  author        = {Xu, Junzhong and Does, Mark D. and Gore, John C.},
  title         = {Quantitative characterization of tissue microstructure with temporal diffusion spectroscopy},
  journal       = {Journal of Magnetic Resonance},
  year          = {2009},
  volume        = {200},
  number        = {2},
  pages         = {189--197},
  %ISSN          = {10907807},
  abstract      = {The signals recorded by diffusion-weighted magnetic resonance imaging (DWI) are dependent on the micro-structural properties of biological tissues, so it is possible to obtain quantitative structural information non-invasively from such measurements. Oscillating gradient spin echo (OGSE) methods have the ability to probe the behavior of water diffusion over different time scales and the potential to detect variations in intracellular structure. To assist in the interpretation of OGSE data, analytical expressions have been derived for diffusion-weighted signals with OGSE methods for restricted diffusion in some typical structures, including parallel planes, cylinders and spheres, using the theory of temporal diffusion spectroscopy. These analytical predictions have been confirmed with computer simulations. These expressions suggest how OGSE signals from biological tissues should be analyzed to characterize tissue microstructure, including how to estimate cell nuclear sizes. This approach provides a model to interpret diffusion data obtained from OGSE measurements that can be used for applications such as monitoring tumor response to treatment in vivo. ?? 2009 Elsevier Inc. All rights reserved.},
  %archiveprefix = {arXiv},
  %arxivid       = {NIHMS150003},
  %doi           = {10.1016/j.jmr.2009.06.022},
  %eprint        = {NIHMS150003},
  file          = {:C$\backslash$:/Users/nmnau/AppData/Local/Mendeley Ltd./Mendeley Desktop/Downloaded/Xu, Does, Gore - 2009 - Quantitative characterization of tissue microstructure with temporal diffusion spectroscopy.pdf:pdf},
  isbn          = {1557-8968 (Electronic)},
  keywords      = {Diffusion time,Modeling,Oscillating gradient,Simulation,Temporal diffusion spectroscopy},
  pmid          = {19616979},
  publisher     = {Elsevier Inc.},
  %url           = {http://dx.%doi.org/10.1016/j.jmr.2009.06.022},
}

@Article{Drobnjak2010,
  author    = {Drobnjak, Ivana and Siow, Bernard and Alexander, Daniel C},
  title     = {Optimizing gradient waveforms for microstructure sensitivity in diffusion-weighted {MR}},
  journal   = {Journal of Magnetic Resonance},
  year      = {2010},
  volume    = {206},
  number    = {1},
  pages     = {41--51},
  publisher = {Elsevier},
}

@Article{Sobol1967-sequence,
  author    = {Sobol, Ilya M},
  title     = {On the distribution of points in a cube and the approximate evaluation of integrals},
  journal   = {Zhurnal Vychislitel'noi Matematiki i Matematicheskoi Fiziki},
  year      = {1967},
  volume    = {7},
  number    = {4},
  pages     = {784--802},
  publisher = {Russian Academy of Sciences, Branch of Mathematical Sciences},
}

@Article{Grebenkov2010,
  author    = {Grebenkov, Denis S},
  title     = {Use, misuse, and abuse of apparent diffusion coefficients},
  journal   = {Concepts in Magnetic Resonance Part A: An Educational Journal},
  year      = {2010},
  volume    = {36},
  number    = {1},
  pages     = {24--35},
  publisher = {Wiley Online Library},
}

@InProceedings{fanDTasia,
  author       = {Barmpoutis, Angelos and Vemuri, Baba C},
  title        = {A unified framework for estimating diffusion tensors of any order with symmetric positive-definite constraints},
  booktitle    = {Biomedical Imaging: From Nano to Macro, 2010 IEEE International Symposium on},
  year         = {2010},
  pages        = {1385--1388},
  organization = {IEEE},
}

@Article{hwang2003image,
  author    = {Hwang, Scott N and Chin, Chih-Liang and Wehrli, Felix W and Hackney, David B},
  title     = {An image-based finite difference model for simulating restricted diffusion},
  journal   = {Magnetic Resonance in Medicine},
  year      = {2003},
  volume    = {50},
  number    = {2},
  pages     = {373--382},
  publisher = {Wiley Online Library},
}

@Article{beltrachini2015parametric,
  author    = {Beltrachini, Leandro and Taylor, Zeike A and Frangi, Alejandro F},
  title     = {A parametric finite element solution of the generalised {Bloch-Torrey} equation for arbitrary domains},
  journal   = {Journal of Magnetic Resonance},
  year      = {2015},
  volume    = {259},
  pages     = {126--134},
  publisher = {Elsevier},
}

@Article{torrey1956bloch,
  author    = {Torrey, Henry C},
  title     = {{Bloch} equations with diffusion terms},
  journal   = {Physical Review},
  year      = {1956},
  volume    = {104},
  number    = {3},
  pages     = {563},
  publisher = {APS},
}

@Article{purslow2002structure,
  author    = {Purslow, Peter P},
  title     = {The structure and functional significance of variations in the connective tissue within muscle},
  journal   = {Comparative Biochemistry and Physiology Part A: Molecular \& Integrative Physiology},
  year      = {2002},
  volume    = {133},
  number    = {4},
  pages     = {947--966},
  publisher = {Elsevier},
}

@Article{Kim2005,
  author   = {Kim, Sungheon and Chi-Fishman, Gloria and Barnett, Alan S and Pierpaoli, Carlo},
  title    = {Dependence on diffusion time of apparent diffusion tensor of ex vivo calf tongue and heart},
  journal  = {Magnetic Resonance in Medicine},
  year     = {2005},
  volume   = {54},
  number   = {6},
  pages    = {1387--1396},
  %ISSN     = {07403194},
  abstract = {The time dependence of the apparent diffusion tensor of ex vivo calf heart and tongue was measured for diffusion times (tau(d)) between 32 and 810 ms. The results showed evidence of restricted diffusion in the muscle tissues of both organs. In regions where the myofibers are parallel, the largest eigenvalue (lambda(1)) of the diffusion tensor remained the same for all diffusion times measured, while the other eigenvalues (lambda(2), lambda(3)) decreased by 29-36{\%} between tau(d) = 32 ms and tau(d) = 400 ms. In regions where the fibers cross, the lambda(1) also changed, decreasing by 17{\%} between tau(d) = 32 ms and tau(d) = 400 ms. The restricting compartment size and volume fraction were effectively estimated by fitting the time courses of the eigenvalues to a model consisting of a nonrestricted compartment and a cylindrically restricted compartment. To our knowledge, this study is the first demonstrating diffusion time dependence of measured water diffusion tensor in muscular tissue. With improvement in scanning technology, future studies may permit noninvasive, in vivo detection of changes in muscle myoarchitecture due to disease, treatment, and exercise.},
  %doi      = {10.1002/mrm.20676},
  file     = {:C$\backslash$:/Users/Noel/AppData/Local/Mendeley Ltd./Mendeley Desktop/Downloaded/Kim et al. - Unknown - Dependence on Diffusion Time of Apparent Diffusion Tensor of Ex Vivo Calf Tongue and Heart.pdf:pdf},
  isbn     = {0740-3194 (Print)$\backslash$n0740-3194 (Linking)},
  keywords = {DTI,Diffusion time,Heart,Restricted water diffusion,Tongue},
  pmid     = {16265644},
}

@Article{Karampinos2009,
  author    = {Karampinos, Dimitrios C. and King, Kevin F. and Sutton, Bradley P. and Georgiadis, John G.},
  title     = {Myofiber ellipticity as an explanation for transverse asymmetry of skeletal muscle diffusion {MRI} in vivo signal},
  journal   = {Annals of Biomedical Engineering},
  year      = {2009},
  volume    = {37},
  number    = {12},
  pages     = {2532--2546},
  %ISSN      = {00906964},
  abstract  = {Due to its unique non-invasive microstructure probing capabilities, diffusion tensor imaging (DTI) constitutes a valuable tool in the study of fiber orientation in skeletal muscles. By implementing a DTI sequence with judiciously chosen directional encoding to quantify in vivo the microarchitectural properties in the calf muscles of three healthy volunteers at rest, we report that the secondary eigenvalue is significantly higher than the tertiary eigenvalue, a phenomenon corroborated by prior DTI findings. Toward a physics-based explanation of this phenomenon, we propose a composite medium model that accounts for water diffusion in the space within the muscle fiber and the extracellular space. The muscle fibers are abstracted as cylinders of infinite length with an elliptical cross section, the latter closely approximating microstructural features well documented in prior histological studies of excised muscle. The range of values of fiber ellipticity predicted by our model agrees with these studies, and the spatial orientation of the cross-sectional ellipses is consistent with local muscle strain fields and the putative direction of lateral transmission of stress between fibers in certain regions in three antigravity muscles (Tibialis Anterior, Soleus, and Gastrocnemius), as well as independent measurements of deformation in active calf muscles. As a metric, fiber cross-sectional ellipticity may be useful for quantifying morphological changes in skeletal muscle fibers with aging, hypertrophy, or sarcopenia.},
  %doi       = {10.1007/s10439-009-9783-1},
  file      = {:C$\backslash$:/Users/Noel/AppData/Local/Mendeley Ltd./Mendeley Desktop/Downloaded/Karampinos et al. - 2009 - Myofiber Ellipticity as an Explanation for Transverse Asymmetry of Skeletal Muscle Diffusion {MRI} In Vivo S(2).pdf:pdf},
  isbn      = {0090-6964},
  keywords  = {Diffusion tensor imaging,Gastrocnemius,Myocytes,Myofiber ellipticity,Skeletal muscle,Soleus},
  pmid      = {19763830},
  publisher = {Springer US},
  %url       = {http://link.springer.com/10.1007/s10439-009-9783-1},
}

@Article{Saotome2006,
  author    = {Saotome, Takako and Sekino, Masaki and Eto, Fumio and Ueno, Shoogo},
  title     = {Evaluation of diffusional anisotropy and microscopic structure in skeletal muscles using magnetic resonance},
  journal   = {Magnetic Resonance Imaging},
  year      = {2006},
  volume    = {24},
  number    = {1},
  pages     = {19--25},
  %ISSN      = {0730725X},
  abstract  = {The pulsed-gradient spin-echo ({PGSE}) nuclear magnetic resonance ({NMR}) method is used for detecting the diffusion of water molecules in biological tissues. Because tissues generally have diffusional anisotropy, their diffusion properties are denoted by a tensor. In this study, we evaluated the diffusional anisotropy and microscopic structure in atrophied skeletal muscles using the {PGSE} {NMR} method. The left sciatic nerve was severed in twelve 9-week-old rats. Neurotomy caused neurogenic muscular atrophy at the left gastrocnemius. At 2, 4 and 8 weeks after neurotomy, magnetic resonance signals were selectively acquired from a 2×2×2 mm3voxel, which was located on the left gastrocnemius. The diffusion tensor, the mean diffusivity (MD) and the fractional anisotropy (FA) were calculated from the signals. A theoretical model of the diffusion in muscles was derived from Tanner's equation. The muscle fiber diameter was estimated by fitting the model to the measured signals. The measurements were also performed for normal rats as controls. No significant difference was found in the MD and the estimated intracellular diffusion coefficient between the control group and the denervated group. The denervated group had significantly higher FA compared with the control group (P{\textless}.05). The estimated muscle fiber diameter of the denervated group was significantly smaller than the estimated value of the control group (P{\textless}.05). These differences were found at 8 weeks after neurotomy. The proposed method is effective for evaluating changes in the microscopic structure of skeletal muscles. {\textcopyright} 2006 Elsevier Inc. All rights reserved.},
  %doi       = {10.1016/j.{MRI}.2005.09.009},
  file      = {:C$\backslash$:/Users/Noel/AppData/Local/Mendeley Ltd./Mendeley Desktop/Downloaded/Saotome et al. - 2006 - Evaluation of diffusional anisotropy and microscopic structure in skeletal muscles using magnetic resonance.pdf:pdf},
  isbn      = {0730-725X},
  keywords  = {Diffusion tensor,Muscle atrophy,Neurotomy,Nuclear magnetic resonance,Pulsed-gradient spin-echo},
  pmid      = {16410174},
  publisher = {Elsevier},
  %url       = {https://www.sciencedirect.com/science/article/pii/S0730725X05002973},
}

@Article{Galban2004,
  author    = {Galb{\'{a}}n, Craig J. and Maderwald, Stefan and Uffmann, Kai and de Greiff, Armin and Ladd, Mark E.},
  title     = {Diffusive sensitivity to muscle architecture: a magnetic resonance diffusion tensor imaging study of the human calf},
  journal   = {European Journal of Applied Physiology},
  year      = {2004},
  volume    = {93},
  number    = {3},
  pages     = {253--262},
  %ISSN      = {1439-6319},
  %doi       = {10.1007/s00421-004-1186-2},
  file      = {:C$\backslash$:/Users/Noel/AppData/Local/Mendeley Ltd./Mendeley Desktop/Downloaded/Galb{\'{a}}n et al. - 2004 - Diffusive sensitivity to muscle architecture a magnetic resonance diffusion tensor imaging study of the human (2).pdf:pdf},
  publisher = {Springer-Verlag},
  %url       = {http://link.springer.com/10.1007/s00421-004-1186-2},
}

@Article{Galban2005,
  author    = {Galb{\'{a}}n, Craig J. and Maderwald, Stefan and Uffmann, Kai and Ladd, Mark E.},
  title     = {A diffusion tensor imaging analysis of gender differences in water diffusivity within human skeletal muscle},
  journal   = {{NMR} in Biomedicine},
  year      = {2005},
  volume    = {18},
  number    = {8},
  pages     = {489--498},
  %ISSN      = {0952-3480},
  %doi       = {10.1002/nbm.975},
  file      = {:C$\backslash$:/Users/Noel/AppData/Local/Mendeley Ltd./Mendeley Desktop/Downloaded/Galb{\'{a}}n et al. - 2005 - A diffusion tensor imaging analysis of gender differences in water diffusivity within human skeletal muscle.pdf:pdf},
  keywords  = {DTI,gender,mathematical modeling,skeletal muscle},
  publisher = {Wiley-Blackwell},
  %url       = {http://%doi.wiley.com/10.1002/nbm.975},
}

@Article{Berry2018,
  author    = {Berry, David B. and Regner, Benjamin and Galinsky, Vitaly and Ward, Samuel R. and Frank, Lawrence R.},
  title     = {Relationships between tissue microstructure and the diffusion tensor in simulated skeletal muscle},
  journal   = {Magnetic Resonance in Medicine},
  year      = {2018},
  volume    = {80},
  number    = {1},
  pages     = {317--329},
  %ISSN      = {15222594},
  abstract  = {Purpose: To establish a series of relationships defining how mus-cle microstructure and diffusion tensor imaging (DTI) are related. Methods: The relationship among key microstructural features of skeletal muscle (fiber size, fibrosis, edema, and permeabil-ity) and the diffusion tensor were systematically simulated over physiologically relevant dimensions individually, and in combi-nation, using a numerical simulation application. Stepwise mul-tiple regression was used to identify which microstructural features of muscle significantly predict the diffusion tensor using single-echo and multi-echo DTI pulse sequences. Simu-lations were also performed in models with histology-informed geometry to investigate the relationship between fiber size and the diffusion tensor in models with real muscle geometry. Results: Fiber size is the strongest predictor of l2, l3, mean diffu-sivity, and fractional anisotropy in skeletal muscle, accounting for approximately 40{\%} of the variance in the diffusion model when cal-culated with single-echo DTI. This increased to approximately 70{\%} when diffusion measures were calculated from the short T 2 compo-nent of the multi-echo DTI sequence. This nonlinear relationship begins to plateau in fibers with greater than 60-mm diameter. Conclusions: As the normal fiber size of a human muscle fiber is 40 to 60 mm, this suggests that DTI is a sensitive tool to monitor muscle atrophy, but may be limited in measurements of muscle with larger fibers. Magn Reson Med 000:000–000, 2017. V C 2017 International Society for Magnetic Resonance in Medicine.},
  %doi       = {10.1002/mrm.26993},
  file      = {:C$\backslash$:/Users/Noel/AppData/Local/Mendeley Ltd./Mendeley Desktop/Downloaded/Berry et al. - 2018 - Relationships between tissue microstructure and the diffusion tensor in simulated skeletal muscle.pdf:pdf},
  keywords  = {DTI,diffusion,multi-echo DTI,muscle microstructure,simulation,skeletal muscle},
  pmid      = {29090480},
  publisher = {Wiley-Blackwell},
  %url       = {http://%doi.wiley.com/10.1002/mrm.26993},
}

@Article{Winters2018,
  author    = {Winters, Kerryanne V. and Reynaud, Olivier and Novikov, Dmitry S. and Fieremans, Els and Kim, Sungheon Gene},
  title     = {Quantifying myofiber integrity using diffusion {MRI} and random permeable barrier modeling in skeletal muscle growth and {Duchenne} muscular dystrophy model in mice},
  journal   = {Magnetic Resonance in Medicine},
  year      = {2018},
  %ISSN      = {15222594},
  abstract  = {{\textcopyright} 2018 International Society for Magnetic Resonance in Medicine. Purpose: To measure the microstructural changes during skeletal muscle growth and progressive pathologies using the random permeable model with diffusion {MRI}, and compare findings to conventional imaging modalities such as three-point Dixon and T 2 imaging. Methods: In vivo and ex vivo DTI experiments with multiple diffusion times (20-700 ms) were completed on wild-type (n=22) and muscle-dystrophic mdx mice (n=8) at various developmental time points. The DTI data were analyzed with the random permeable model framework that provides estimates of the unrestricted diffusion coefficient (D 0 ), membrane surface-to-volume ratio (S/V), and membrane permeability ($\kappa$). In addition, the {MRI} experiments included conventional measures, such as tissue fat fractions and T 2 relaxation. Results: During normal muscle growth between week 4 and week 13, the in vivo S/V, fractional anisotropy, and fat fraction correlated positively with age ($\rho$=0.638, 0.664, and 0.686, respectively), whereas T 2 correlated negatively ($\rho$=-0.847). In mdx mice, all DTI random permeable model parameters and fat fraction had significant positive correlation with age, whereas fractional anisotropy and T 2 did not have significant correlation with age. Histological measurements of the perimeter-to-area ratio served as a proxy for the model-derived S/V in the cylindrical myofiber geometry, and had a significant correlation with the ex vivo S/V (r=0.71) as well as the in vivo S/V (r=0.56). Conclusion: The present study demonstrates that DTI at multiple diffusion times with the random permeable model analysis allows for noninvasively quantifying muscle fiber microstructural changes during both normal muscle growth and disease progression. Future studies can apply our technique to evaluate current and potential treatments to muscle myopathies.},
  %doi       = {10.1002/mrm.27188},
  file      = {:C$\backslash$:/Users/Noel/AppData/Local/Mendeley Ltd./Mendeley Desktop/Downloaded/Winters et al. - 2018 - Quantifying myofiber integrity using diffusion {MRI} and random permeable barrier modeling in skeletal muscle grow.pdf:pdf},
  isbn      = {1522-2594 (Electronic) 0740-3194 (Linking)},
  keywords  = {Diffusion tensor imaging,Diffusion time,Random permeable barrier model,Sarcolemma permeability,Surface-to-volume ratio},
  pmid      = {29577406},
  publisher = {Wiley-Blackwell},
  %url       = {http://%doi.wiley.com/10.1002/mrm.27188},
}

@Article{Porcari2018,
  author    = {Porcari, Paola and Hall, Matt G. and Clark, Chris A. and Greally, Elizabeth and Straub, Volker and Blamire, Andrew M.},
  title     = {The effects of ageing on mouse muscle microstructure: a comparative study of time-dependent diffusion {MRI} and histological assessment},
  journal   = {{NMR} in Biomedicine},
  year      = {2018},
  volume    = {31},
  number    = {3},
  pages     = {e3881},
  %ISSN      = {10991492},
  abstract  = {{\textcopyright} 2018 John Wiley {\&} Sons, Ltd. The investigation of age-related changes in muscle microstructure between developmental and healthy adult mice may help us to understand the clinical features of early-onset muscle diseases, such as Duchenne muscular dystrophy. We investigated the evolution of mouse hind-limb muscle microstructure using diffusion imaging of in vivo and in vitro samples from both actively growing and mature mice. Mean apparent diffusion coefficients ({ADC}s) of the gastrocnemius and tibialis anterior muscles were determined as a function of diffusion time ($\Delta$), age (7.5, 22 and 44 weeks) and diffusion gradient direction, applied parallel or transverse to the principal axis of the muscle fibres. We investigated a wide range of diffusion times with the goal of probing a range of diffusion lengths characteristic of muscle microstructure. We compared the diffusion time-dependent {ADC} of hind-limb muscles with histology. {ADC} was found to vary as a function of diffusion time in muscles at all stages of maturation. Muscle water diffusivity was higher in younger (7.5 weeks) than in adult (22 and 44 weeks) mice, whereas no differences were observed between the older ages. In vitro data showed the same diffusivity pattern as in vivo data. The highlighted differences in diffusion properties between young and mature muscles suggested differences in underlying muscle microstructure, which were confirmed by histological assessment. In particular, although diffusion was more restricted in older muscle, muscle fibre size increased significantly from young to adult age. The extracellular space decreased with age by only {\~{}}1{\%}. This suggests that the observed diffusivity differences between young and adult muscles may be caused by increased membrane permeability in younger muscle associated with properties of the sarcolemma.},
  %doi       = {10.1002/nbm.3881},
  file      = {:C$\backslash$:/Users/Noel/AppData/Local/Mendeley Ltd./Mendeley Desktop/Downloaded/Porcari et al. - 2018 - The effects of ageing on mouse muscle microstructure a comparative study of time-dependent diffusion {MRI} and his.pdf:pdf},
  keywords  = {apparent diffusion coefficient,diffusion {MRI},diffusion-weighted imaging,hind-limb muscles,histology,maturation,mouse,time-dependent diffusion {MRI}},
  pmid      = {29315904},
  publisher = {Wiley-Blackwell},
  %url       = {http://%doi.wiley.com/10.1002/nbm.3881},
}

@Article{Hall2017,
  author    = {Hall, Matt G. and Clark, Chris A.},
  title     = {Diffusion in hierarchical systems: A simulation study in models of healthy and diseased muscle tissue},
  journal   = {Magnetic Resonance in Medicine},
  year      = {2017},
  volume    = {78},
  number    = {3},
  pages     = {1187--1198},
  %ISSN      = {15222594},
  %doi       = {10.1002/mrm.26469},
  file      = {:C$\backslash$:/Users/Noel/AppData/Local/Mendeley Ltd./Mendeley Desktop/Downloaded/Hall, Clark - 2017 - Diffusion in hierarchical systems A simulation study in models of healthy and diseased muscle tissue.pdf:pdf},
  isbn      = {1522-2594 (Electronic)
0740-3194 (Linking)},
  keywords  = {Duchenne muscular dystrophy,{Monte}-{Carlo} simulation,diffusion {MRI},entropy,muscle,pathology modeling},
  pmid      = {27667781},
  publisher = {Wiley-Blackwell},
  %url       = {http://%doi.wiley.com/10.1002/mrm.26469},
}

@Article{Laghi2017,
  author    = {Laghi, Luca and Venturi, Luca and Dellarosa, Nicol{\`{o}} and Petracci, Massimiliano},
  title     = {Water diffusion to assess meat microstructure},
  journal   = {Food Chemistry},
  year      = {2017},
  volume    = {236},
  pages     = {15--20},
  %ISSN      = {18737072},
  abstract  = {In the quest for setting up rapid methods to evaluate water retention ability of meat microstructures, time domain nuclear magnetic resonance (TD-{NMR}) has gained a prominent role, due to the possibility to observe water located outside the myofibrils, easily lost upon storage or cooking. Diffusion weighted signals could be used to monitor the shape and dimension of the pores in which water is confined, thus boosting the information offered by TD-{NMR}. The work outlines a parsimonious model to describe relative abundance and diffusion coefficient of intra and extra myofibrillar water populations, exchange rate between them, diameter of the myofibrillar cells. To test our model, we registered diffusion and {T2} weighted {NMR} signals at 20 MHz on fresh meat from pectoralis major muscle of 100 days old female turkey. We then purposely altered water distribution and myofibrils shape by means of freezing. The model predicted nicely the consequences of the imposed modifications.},
  %doi       = {10.1016/j.foodchem.2016.12.043},
  file      = {:C$\backslash$:/Users/Noel/AppData/Local/Mendeley Ltd./Mendeley Desktop/Downloaded/Laghi et al. - 2017 - Water diffusion to assess meat microstructure.pdf:pdf},
  keywords  = {Meat thawing,TD-{NMR},Water compartmentalization,Water diffusion},
  publisher = {Elsevier},
  %url       = {https://www.sciencedirect.com/science/article/pii/S0308814616320520},
}

@Article{Damon2008,
  author    = {Damon, Bruce M.},
  title     = {Effects of image noise in muscle diffusion tensor {(DT)-MRI} assessed using numerical simulations},
  journal   = {Magnetic Resonance in Medicine},
  year      = {2008},
  volume    = {60},
  number    = {4},
  pages     = {934--944},
  %ISSN      = {07403194},
  abstract  = {Diffusion tensor (DT)-{MRI} studies of skeletal muscle provide information about muscle architecture, microstructure, and damage. However, the effects of noise, the diffusion weighting (b)-value, and partial volume artifacts on the estimation of the diffusion tensor (D) are unknown. This study investigated these issues using {Monte} {Carlo} simulations of 3 x 9 voxel regions of interest (ROIs) containing muscle, adipose tissue, and intermediate degrees of muscle volume fractions (f(M)). A total of 1000 simulations were performed for each of eight b-values and 11 SNR levels. The dependencies of the eigenvalues (lambda(1-3)), mean diffusivity (lambda), and fractional anisotropy (FA), and the angular deviation of the first eigenvector from its true value (alpha) were observed. For moderate b-values (b = 435-725 s/mm(2)) and f(M) = 1, an accuracy of 5{\%} was obtained for lambda(1-3), lambda, and FA with an SNR of 25. An accuracy of 1{\%} was obtained for lambda(1-3), lambda, and FA with f(M) = 1 and SNR = 50. For regions with f(M) = 8/9, 5{\%} accuracy was obtained with SNR = 40. For alpha, SNRs of {\textgreater}or=25 and {\textgreater}or=45 were required for +/-4.5 degrees uncertainty with f(M) = 1 and f(M) = 0.5, respectively; SNR {\textgreater}or= 60 was required for +/-9 degrees uncertainty in single muscle voxels. These findings may influence the design and interpretation of DT-{MRI} studies of muscle microstructure, damage, and architecture.},
  %doi       = {10.1002/mrm.21707},
  file      = {:C$\backslash$:/Users/Noel/AppData/Local/Mendeley Ltd./Mendeley Desktop/Downloaded/Damon - 2008 - Effects of image noise in muscle diffusion tensor (DT)-{MRI} assessed using numerical simulations.pdf:pdf},
  isbn      = {0740-3194},
  keywords  = {Accuracy,DTI,Diffusion,Fiber tracking,Noise,SNR,Tensor},
  pmid      = {18816814},
  publisher = {Wiley-Blackwell},
  %url       = {http://%doi.wiley.com/10.1002/mrm.21707},
}

@Article{Fieremans2017,
  author   = {Fieremans, Els and Lemberskiy, Gregory and Veraart, Jelle and Sigmund, Eric E. and Gyftopoulos, Soterios and Novikov, Dmitry S.},
  title    = {In vivo measurement of membrane permeability and myofiber size in human muscle using time-dependent diffusion tensor imaging and the random permeable barrier model},
  journal  = {{NMR} in Biomedicine},
  year     = {2017},
  volume   = {30},
  number   = {3},
  pages    = {e3612},
  %ISSN     = {10991492},
  abstract = {The time dependence of the diffusion coefficient is a hallmark of tissue complexity at the micrometer level. Here we demonstrate how biophysical modeling, combined with a specifically tailored diffusion {MRI} acquisition performing diffusion tensor imaging (DTI) for varying diffusion times, can be used to determine fiber size and membrane permeability of muscle fibers in vivo. We describe the random permeable barrier model (RPBM) and its assumptions, as well as the details of stimulated echo DTI acquisition, signal processing steps, and potential pitfalls. We illustrate the RPBM method on a few pilot examples involving human subjects (previously published as well as new), such as revealing myofiber size derived from RPBM increase after training in a calf muscle, and size decrease with atrophy in shoulder rotator cuff muscle. Finally, we comment on the potential clinical relevance of our results. Copyright (c) 2016 John Wiley {\&} Sons, Ltd.},
  %doi      = {10.1002/nbm.3612},
  file     = {:C$\backslash$:/Users/Noel/AppData/Local/Mendeley Ltd./Mendeley Desktop/Downloaded/Fieremans et al. - 2017 - iIn vivoi measurement of membrane permeability and myofiber size in human muscle using time-dependent diffusio.pdf:pdf},
  isbn     = {0952-3480},
  keywords = {DTI,exercise,random permeable barrier model,rotator cuff muscle,skeletal muscle,stimulated echo,time-dependent diffusion},
  pmid     = {27717099},
  %url      = {http://%doi.wiley.com/10.1002/nbm.3612},
}

@Article{Bates2017,
  author          = {Bates, Joanne and Teh, Irvin and McClymont, Darryl and Kohl, Peter and Schneider, J{\"{u}}rgen E. and Grau, Vicente},
  title           = {{Monte Carlo} Simulations of Diffusion Weighted {MRI} in Myocardium: Validation and Sensitivity Analysis},
  journal         = {IEEE Transactions on Medical Imaging},
  year            = {2017},
  volume          = {36},
  number          = {6},
  pages           = {1316--1325},
  %ISSN            = {1558254X},
  %doi             = {10.1109/TMI.2017.2679809},
  isbn            = {0278-0062},
  keywords        = {{MRI},diffusion weighted imaging,heart,tissue modelling},
  mendeley-groups = {ALL d{MRI} PAPERS,Muscle/{MRI} and Muscle/d{MRI}},
  %url             = {http://ieeexplore.ieee.org/document/7875442/},
}

@article{Correa-de-Araujo2017a,
  author    = {Correa-de-Araujo, Rosaly and Harris-Love, Michael O. and Miljkovic, Iva and Fragala, Maren S. and Anthony, Brian W. and Manini, Todd M.},
  title     = {The Need for Standardized Assessment of Muscle Quality in Skeletal Muscle Function Deficit and Other Aging-Related Muscle Dysfunctions: A Symposium Report},
  year      = {2017},
  abstract  = {A growing body of scientific literature suggests that not only changes in skeletal muscle mass, but also other factors underpinning muscle quality, play a role in the decline in skeletal muscle function and impaired mobility associated with aging. A symposium on muscle quality and the need for standardized assessment was held on April 28, 2016 at the International Conference on Frailty and Sarcopenia Research in Philadelphia, Pennsylvania. The purpose of this symposium was to provide a venue for basic science and clinical researchers and expert clinicians to discuss muscle quality in the context of skeletal muscle function deficit and other aging-related muscle dysfunctions. The present article provides an expanded introduction concerning the emerging definitions of muscle quality and a potential framework for scientific inquiry within the field. Changes in muscle tissue composition, based on excessive levels of inter- and intra-muscular adipose tissue and intramyocellular lipids, have been found to adversely impact metabolism and peak force generation. However, methods to easily and rapidly assess muscle tissue composition in multiple clinical settings and with minimal patient burden are needed. Diagnostic ultrasound and other assessment methods continue to be developed for characterizing muscle pathology, and enhanced sonography using sensors to provide user feedback and improve reliability is currently the subject of ongoing investigation and development. In addition, measures of relative muscle force such as specific force or grip strength adjusted for body size have been proposed as methods to assess changes in muscle quality. Furthermore, performance-based assessments of muscle power via timed tests of function and body size estimates, are associated with lower extremity muscle strength may be responsive to age-related changes in muscle quality. Future aims include reaching consensus on the definition and standardized assessments of muscle quality, and providing recommendations to address critical clinical and technology research gaps within the field.},
  journal = {Frontiers in Physiology},
  %doi       = {10.3389/fphys.2017.00087},
  file      = {:C$\backslash$:/Users/Noel/AppData/Local/Mendeley Ltd./Mendeley Desktop/Downloaded/Correa-de-Araujo et al. - 2017 - The Need for Standardized Assessment of Muscle Quality in Skeletal Muscle Function Deficit and Other Ag.pdf:pdf},
  isbn      = {1664-042X (Print) 1664-042X (Linking)},
  %ISSN      = {1664042X},
  keywords  = {Imaging,Muscle power,Muscle quality,Muscle strength,Myosteatosis,Sarcopenia,Skeletal muscle function deficit},
  pages     = {87},
  pmid      = {28261109},
  publisher = {Frontiers},
  %url       = {http://journal.frontiersin.org/article/10.3389/fphys.2017.00087/full},
  volume    = {8},
}

@Article{Villareal2004,
  author    = {Villareal, Dennis T. and Banks, Marian and Siener, Catherine and Sinacore, David R. and Klein, Samuel},
  title     = {Physical frailty and body composition in obese elderly men and women},
  journal   = {Obesity Research},
  year      = {2004},
  volume    = {12},
  number    = {6},
  pages     = {913--920},
  %ISSN      = {10717323},
  abstract  = {To evaluate the prevalence of frailty and interrelationships among body composition, physical function, and quality of life in community-dwelling obese elderly (OE) persons.},
  %doi       = {10.1038/oby.2004.111},
  file      = {:C$\backslash$:/Users/Noel/AppData/Local/Mendeley Ltd./Mendeley Desktop/Downloaded/Villareal et al. - 2004 - Physical Frailty and Body Composition in Obese Elderly Men and Women(2).pdf:pdf},
  isbn      = {1071-7323},
  keywords  = {Body composition,Elderly,Muscle strength,Physical frailty,Quality of life},
  pmid      = {15229329},
  publisher = {Wiley-Blackwell},
  %url       = {http://%doi.wiley.com/10.1038/oby.2004.111},
}

@Article{Wokke2014,
  author    = {Wokke, B. H. and van den Bergen, J. C. and Versluis, M. J. and Niks, E. H. and Milles, J. and Webb, A. G. and van Zwet, E. W. and Aartsma-Rus, A. and Verschuuren, J. J. and Kan, H. E.},
  title     = {Quantitative {MRI} and strength measurements in the assessment of muscle quality in {Duchenne} muscular dystrophy},
  journal   = {Neuromuscular Disorders},
  year      = {2014},
  volume    = {24},
  number    = {5},
  pages     = {409--416},
  %ISSN      = {18732364},
  abstract  = {The purpose of this study was to assess leg muscle quality and give a detailed description of leg muscle involvement in a series of Duchenne muscular dystrophy patients using quantitative {MRI} and strength measurements. Fatty infiltration, as well as total and contractile (not fatty infiltrated) cross sectional areas of various leg muscles were determined in 16 Duchenne patients and 11 controls (aged 8-15). To determine specific muscle strength, four leg muscle groups (quadriceps femoris, hamstrings, anterior tibialis and triceps surae) were measured and related to the amount of contractile tissue. In patients, the quadriceps femoris showed decreased total and contractile cross sectional area, attributable to muscle atrophy. The total, but not the contractile, cross sectional area of the triceps surae was increased in patients, corresponding to hypertrophy. Specific strength decreased in all four muscle groups of Duchenne patients, indicating reduced muscle quality. This suggests that muscle hypertrophy and fatty infiltration are two distinct pathological processes, differing between muscle groups. Additionally, the quality of remaining muscle fibers is severely reduced in the legs of Duchenne patients. The combination of quantitative {MRI} and quantitative muscle testing could be a valuable outcome parameter in longitudinal studies and in the follow-up of therapeutic effects. {\textcopyright} 2014 Elsevier B.V.},
  %doi       = {10.1016/j.nmd.2014.01.015},
  file      = {:C$\backslash$:/Users/Noel/AppData/Local/Mendeley Ltd./Mendeley Desktop/Downloaded/Wokke et al. - 2014 - Quantitative {MRI} and strength measurements in the assessment of muscle quality in Duchenne muscular dystrophy.pdf:pdf},
  isbn      = {0960-8966},
  keywords  = {Duchenne muscular dystrophy,Muscle quality,Quantitative {MRI}},
  pmid      = {24613733},
  publisher = {Elsevier},
  %url       = {https://www.sciencedirect.com/science/article/pii/S0960896614000480},
}

@article{Heymsfield2015,
  author    = {Heymsfield, Steven B. and Gonzalez, M. Cristina and Lu, Jianhua and Jia, Guang and Zheng, Jolene},
  title     = {Skeletal muscle mass and quality: Evolution of modern measurement concepts in the context of sarcopenia},
  journal   = {Proceedings of the Nutrition Society},
  year      = {2015},
  volume    = {74},
  number    = {4},
  pages     = {355--366},
  publisher = {Cambridge University Press},
  abstract  = {The first reports of accurate skeletal muscle mass measurement in human subjects appeared at about the same time as introduction of the sarcopenia concept in the late 1980s. Since then these methods, computed tomography and {MRI}, have been used to gain insights into older (i.e. anthropometry and urinary markers) and more recently developed and refined methods (ultrasound, bioimpedance analysis and dual-energy X-ray absorptiometry) of quantifying regional and total body skeletal muscle mass. The objective of this review is to describe the evolution of these methods and their continued development in the context of sarcopenia evaluation and treatment. Advances in these technologies are described with a focus on additional quantifiable measures that relate to muscle composition and ‘quality'. The integration of these collective evaluations with strength and physical performance indices is highlighted with linkages to evaluation of sarcopenia and the spectrum of related disorders such as sarcopenic obesity, cachexia and frailty. Our findings show that currently available methods and those in development are capable of non-invasively extending measures from solely ‘mass' to quality evaluations that promise to close the gaps now recognised between skeletal muscle mass and muscle function, morbidity and mortality. As the largest tissue compartment in most adults, skeletal muscle mass and aspects of muscle composition can now be evaluated by a wide array of technologies that provide important new research and clinical opportunities aligned with the growing interest in the spectrum of conditions associated with sarcopenia.},
  %doi       = {10.1017/S0029665115000129},
  file      = {:C$\backslash$:/Users/Noel/AppData/Local/Mendeley Ltd./Mendeley Desktop/Downloaded/Heymsfield et al. - 2015 - Skeletal muscle mass and quality Evolution of modern measurement concepts in the context of sarcopenia.pdf:pdf},
  isbn      = {0029-6651},
  %ISSN      = {14752719},
  keywords  = {Body composition,Cachexia,Nutritional assessment,Sarcopenia,Sarcopenic obesity},
  pmid      = {25851205},
  %url       = {http://www.journals.cambridge.org/abstract{\_}S0029665115000129},
}

@Article{Roubenoff2001,
  author          = {Roubenoff, Ronenn},
  title           = {Origins and clinical relevance of sarcopenia},
  journal         = {Canadian Journal of Applied Physiology},
  year            = {2001},
  volume          = {26},
  number          = {1},
  pages           = {78--89},
  %ISSN            = {1066-7814},
  abstract        = {Sarcopenia is the loss of muscle mass and strength that occurs with normal aging. Because sarcopenia is not the result of a disease, it is seen in all aged adults. Sarcopenia markedly increases the risk of disability and loss of functional capacity in the elderly. The mechanisms underlying sarcopenia are complex and are reviewed here. It is not clear at this time which factors are most important in determining the severity or rate of development of sarcopenia. While progressive resistance training clearly can reverse and prevent sarcopenia, little is known about the mechanisms by which aged muscle adapts to training, or whether these adaptations reflect reversal of direct pathophysiological processes or compensation by activation of separate pathways from those leading to the deterioration in the first place. As populations in developed countries continue to age, diagnosing, treating, and preventing sarcopenia will be progressively more important to the health and well-being of modern societies.},
  %archiveprefix   = {arXiv},
  %arxivid         = {arXiv:1011.1669v3},
  %doi             = {10.1017/CBO9781107415324.004},
  %eprint          = {arXiv:1011.1669v3},
  file            = {:C$\backslash$:/Users/Noel/AppData/Local/Mendeley Ltd./Mendeley Desktop/Downloaded/Roubenoff - Unknown - Origins and Clinical Relevance of Sarcopenia.pdf:pdf},
  isbn            = {9788578110796},
  keywords        = {icle},
  mendeley-groups = {Muscle/Muscle Quality - Lit Review},
  pmid            = {11291626},
  %url             = {www.nrcresearchpress.com http://www.ncbi.nlm.nih.gov/pubmed/11291626},
}

@Article{Rose2019,
  author    = {Rose, Jan N and Nielles-Vallespin, Sonia and Ferreira, Pedro F and Firmin, David N and Scott, Andrew D and Doorly, Denis J},
  title     = {Novel insights into in-vivo diffusion tensor cardiovascular magnetic resonance using computational modeling and a histology-based virtual microstructure},
  journal   = {Magnetic Resonance in Medicine},
  year      = {2019},
  volume    = {81},
  number    = {4},
  pages     = {2759--2773},
  publisher = {Wiley Online Library},
}

@article{Tanner1970,
	title={Use of the stimulated echo in {NMR} diffusion studies},
	author={Tanner, John E},
	journal={The Journal of Chemical Physics},
	volume={52},
	number={5},
	pages={2523--2526},
	year={1970},
	publisher={AIP}
}
}


\end{document}